\documentclass[12pt,preprint]{aastex}
\slugcomment{Accepted and scheduled for publication  
in {\it the Astrophysical Journal},  for  April 10, 2007, v 659 1 issue}  
\def\lax {\ifmmode{_<\atop^{\sim}}\else{${_<\atop^{\sim}}$}\fi}  
\def\gax {\ifmmode{_>\atop^{\sim}}\else{${_>\atop^{\sim}}$}\fi}  
\def\gtorder{\mathrel{\raise.3ex\hbox{$>$}\mkern-14mu
             \lower0.6ex\hbox{$\sim$}}}

\usepackage{lscape}

\begin{document}

\title{Power spectra of black holes and neutron stars as a probe of  hydrodynamical structure of the source. 
Diffusion theory  and its application to Cyg X-1 and Cyg X-2 X-ray observations}

\author{Lev Titarchuk\altaffilmark{1,2,3}, Nikolai Shaposhnikov\altaffilmark{4} and Vadim Arefiev\altaffilmark{5}  }

\altaffiltext{1}{George Mason University/Center for Earth
Observing and Space Research, Fairfax, VA 22030; and US Naval Research
Laboratory, Code 7655, Washington, DC 20375-5352; ltitarchuk@ssd5.nrl.navy.mil }
\altaffiltext{2}{Dipartimento di Fisica, Universit\'a di Ferrara, via Saragat 1, I--44100, Ferrara, Italy; 
titarchuk@fe.infn.it}
\altaffiltext{3}{Goddard Space Flight Center, NASA, code 661, Greenbelt  
MD 20771; lev@milkyway.gsfc.nasa.gov}
\altaffiltext{4}{Goddard Space Flight Center, NASA/Universities Space Research
Association, code 662, Greenbelt  
MD 20771; nikolai@milkyway.gsfc.nasa.gov}
\altaffiltext{5}{Space Research Institute (IKI), Russian Academy of Science, Profsoyuznaya 84/32, 117997, Moscow, Russia,
gita@hea.iki.rssi.ru}

\begin{abstract}

We present a model of Fourier Power Density Spectrum (PDS) formation in accretion 
powered X-ray binary systems derived from the first principles of the diffusion theory.  
Timing properties of X-ray emission are considered to be a result of diffusive propagation 
of the driving perturbations in a bounded medium. We prove that the integrated power of the resulting PDS, $P_x$  is only a small fraction of the integrated power of the driving oscillations, $P_{dr}$  which is distributed over the disk. Furthermore, we demonstrate that  the power $P_x$ is inversely proportional to the characteristic frequency of the driving oscillations $\nu_{dr}$ which is likely scaled with the frequency of the local gravity waves in the disk (Keplerian frequency).  
Keeping in mind that $\nu_{dr}$ increases towards soft states  leads us to conclude that  the power 
$P_x$   declines towards soft states. This dependence $P_x\propto \nu_{dr}^{-1}$ 
 explains the well-known observational phenomenon that the  power of the X-ray variability decreases when the source evolves to softer states.  
The resulting PDS continuum is a sum of  two components,  a low frequency (LF) component  
which presumably originates  in an extended accretion disk 
and a high frequency (HF) component  which   originates in the innermost part of the source [Compton cloud (CC)].  
The LF  PDS component has a power-law shape with index of $1.0-1.5$ at higher
frequencies (``red'' noise) and a flat spectrum  below  a characteristic (break) frequency (``white'' noise). 
This white-red noise (WRN) continuum
spectrum holds information about the physical parameters of the bounded extended medium, 
diffusion time scale and the dependence law of viscosity vs radius.
 This LF PDS associated with the extended disk  dominates in the soft states of the system, while the HF PDS characteristic of innermost CC component 
is dominant in the low/hard and intermediate states. 
 These PDS LF and HF components  directly correspond to the energy spectrum  components. Namely:
LF WRN is related to thermal emission from an accretion disk, and the HF WRN  to the power-law 
tail, which presents a fraction of the disk emission Comptonized in the Compton cloud.
Hence, a change of PDS  features  correlates with a change of energy spectral features. 
Analyzing the data for a number of sources we find   that 
the PDS is well represented by a sum of 
the WRN CC component and  the WRN extended disk component. 
We apply  our model of the PDS  to a sample of RXTE and EXOSAT timing data from Cyg X-1 and Cyg X-2
 which describes  adequately  the  spectral transitions in these sources. The presented 
PDSs are shown in frequency range from $10^{-8}$ Hz to $10^{2}$ Hz, 
 i.e. in 10 orders of magnitude range. We offer a method to measure an effective Reynolds number, 
(${\rm Re}$) [inverse of the disk viscosity ($\alpha_{\rm SS})-$parameter]
 using the basic power spectrum parameters, PDS index and characteristic  frequencies. 
Our analysis gives ${\rm Re}$ in the range of $8\pm 2.5$, 
 or $\alpha_{\rm SS}\sim({\rm Re})^{-1}=0.14\pm0.04$,  
 for the Cyg X-1 and Cyg X-2 CC configurations which are associated with their HF  PDS component. 
\end{abstract}


\section{Introduction}
In  Astronomy, in general  the basic question is: what one can learn from the observations confronting
 a theory derived from the first principles and the main laws of
Physics.  Particularly,  X-ray Astronomy studies the spectral and timing properties of  X-ray emission sources. During the last three decades a bulk of observational evidence 
emerged, showing that black hole X-ray binaries   evolves through 
a set of spectral states \citep[see][and references therein]{rm06}. The basic properties of 
 the X-ray energy spectra in a particular state are determined by a distribution of
 photons between two major spectral continuum components, i.e. the thermal
 component which comes from an accretion disk and the power law, presumably formed
 by the soft disk photons upscattered in a hot plasma surrounding the disk [
 Compton cloud (CC)].  Specifically, in the low-hard state the energy spectrum of a source is dominated
 by a power law part, while in high-soft states the thermal disk component is dominant. 
The Fourier Power Density Spectrum (PDS) also has a specific shape in each state.
In low-hard state the source emission is highly variable (up to 40\% root-mean-square (rms)
variability) and PDS has a broken power law shape with a flat plateau below the break frequency.
In a less variable (less than 10 \% rms) high-soft state PDS is a power law with index of 1.0-1.5
extending up to an orbital frequency of a binary system \citep[][ hereafter GA06]{ga}
with a cut-off at the higher frequencies. Many efforts have been made to build a consistent
theory of PDS formation in accreting sources (see references below). In this Paper we
present a model which is based on the exact analytical solution of perturbation diffusion
equation in a bounded configuration (i.e. accretion  disk or CC). The model explains
the continuum shape of the observed PDS components as well as additional effects such as 
high frequency cut-off and rms-flux relationship \citep{utt04}.

Black hole (BH) and neutron star (NS) sources are characterized by  hard and soft states and the 
transition between them (intermediate states).
Vignarca et al. (2003), Kalemci (2003),  Titarchuk \& Fiorito (2004), hereafter TF04, Titarchuk \& Shaposhnikov (2005), hereafter TS05, Shaposhnikov \& Titarchuk (2006),
hereafter ST06, found that the spectral and timing
properties are tightly related to each other in a number of BH and NS sources.  Namely, they found  observational evidence for the correlation of
spectral index with low-frequency features: break frequency $\nu_b$ and  quasi-periodic oscillation (QPO)
low frequency $\nu_{L}$. In BHs the photon index $\Gamma$ steadily increases from 1.5
in the low-hard state to values exceeding 2.1 in  soft states. 
In the high/soft state of BH the spectral index-QPO oscillation 
frequency correlation shows a flattening, or  ``saturation'' of the photon
index $\Gamma$ at high values of the QPO frequency $\nu_L$.  This saturation effect was identified as a BH signature. 
TS05 demonstrate that this saturation is not present at least for one NS source. 
They show that for  4U 1728-34  the index $\Gamma$ monotonically increases with $\nu_L$.
ST06  found that Cyg X-1 is a perfect example of the BH source where the BH  index-QPO frequency correlation is  observed with  
clear features of the saturation at high  and low frequencies. 

Furthermore, the values of the break frequency and the QPO frequency are related to the BH mass. 
In principle, one can evaluate the mass of the central object using the index-QPO relation
 because QPO frequencies are inversely proportional to mass (TF04).
The simple scaling of the index-QPO correlation for XTE J1550-564  over the frequency axis gives 
us  the index-QPO correlation for   GRS 1915+105. The shifting factor is
10/12 which gives the relative BH mass in XTE J1550-564 with respect to that in GRS 1915+105. 
However, there is one condition for the method applicability:  the index-QPO relations should be 
self-similar with respect to each other, as it occurs  for GRS 1915+105 and  XTE J1550-564. 

\citet{king04} proposed an explicit physical model for the disc variability, 
consistent with Lyubarskii's general scheme \citep{L97}, hereafter L97, for solving this problem. They
 suggested that local dynamo processes can affect the evolution of the accretion disc by driving angular momentum loss 
 in the form of an outflow (wind or jet). K04 argued that large-scale outflow can only occur when the 
small-scale random processes in neighboring disk annuli give  rise by chance to a coherent large-scale magnetic field. 
 This occurs on much longer time-scales (than that of the 
small-scale random processes), and causes a bright large-amplitude flare as a wide range of disc radii 
 evolve in a coherent fashion.  Most of the time, dynamo action instead produces small-amplitude flickering. 
 
 It is worth noting that in our work we offer  
 a method to measure an effective Reynolds number ${\rm Re}$
 using the basic power spectrum parameters (PDS index and characteristic  frequencies). 
We obtain that the inferred ${\rm Re}$  in the range of $8\pm 2.5$ 
 for the Cyg X-1 and Cyg X-2 CC configurations related to their high frequency  PDS component (see details \S 4.2.2).   
This relatively low value of ${\rm Re}$ means that there is actually a large turbulent structure (comparable with the 
size of the accretion flow configuration).   Thus we come to the same conclusion that in the bounded disk-like 
configuration (CC) there is  a large turbulent structure which can be  a  large-scale magnetic field proposed by K04. 
 
K04 also reproduced power spectra similar to those observed, including a $1/\nu$ power spectrum below a break frequency given 
by  the magnetic alignment time-scale at the inner disc edge. 
 Moreover,  K04 concluded from in their simulation 
of power spectra of the accretion disk that the relation between BH mass and the value of the break frequency is 
less straightforward than often assumed in the literature (cf. ST06).
In this paper we demonstrate that the resulting power spectrum of the BH sources is generally a sum of two components: 
one is presumably related to the innermost part of the disk and another one is related to the oscillations of the extended disk. 
The values of PDS features (break and QPO frequencies) of the former one are scaled with  the BH mass and the break frequency 
value of the latter component is rather scaled with the disk size and they correlate with the binary orbital frequency (GA06).  

\citet{L97} considered small amplitude local fluctuations in the accretion rate at each radius, 
caused by small amplitude variations in the viscosity, and then considers the effect of these fluctuations on the 
accretion rate at the inner disc edge. A linear calculation shows that if the characteristic time-scale of the viscosity 
variations  is everywhere comparable to the viscous (inflow) time-scale, and 
if the amplitude of the variations is independent of radius, then the power spectrum of luminosity fluctuations is a power-law 
$1/\nu$.  If the amplitude of the variations increases with radius, the slope of the power spectrum of the luminosity variations 
is steeper than 1. 
Lyubarskii pointed out that he had no physical model for the cause of such fluctuations. 
In particular, although the obvious candidate cause is the magnetic dynamo, 
the characteristic time-scales for the dynamo are much shorter than the local viscous time-scale.
However, K04 modeled the dynamo as a small-scale stochastic phenomenon, 
 operating on roughly the local dynamical time-scale. 

L97 and its extension by Kotov, Churazov \& Gilfanov  (2001) sought also 
 to explain the spectral-timing properties of 
the X-ray variability of accreting black holes in terms of inward-propagating mass accretion fluctuations produced at 
a broad range of radii. The fluctuations modulate the X-ray emitting region as they move inwards and can produce 
temporal-frequency-dependent lags between energy bands, 
and energy-dependent power PDSs as a result of the different emissivity profiles, 
which may be expected at different X-ray energies. 
Kotov et al. (2001) analytically determined spectral-timing properties by making the simplifying assumption that the perturbation 
introduced into the accretion flow at each radius is a delta function in time and radius. 
 
Recently Mayer \& Pringle (2006), hereafter MP06, extended the K04 model by taking proper account 
of the thermal properties of the disc.  Because the degree of variability in the K04 model depends sensitively on the ratio of disc 
thickness to radius,  $H/R$. MP06 suggested that it was important to follow the time dependence of the local disc structure 
as the variability proceeds. 
MP06 agreed that radial heat advection plays an important role in determining the inner disc structure, and also found  
limit-cycle behavior. 
 
Uttley (2004) and Uttley, McHardy  \& Vaughan (2005)  have pointed out that the existence of a strong linear 
relationship between the amplitude of the X-ray variability and the X-ray flux  - the rms-flux relation - can 
be used a diagnostic to distinguish between various models. They conclude that simplistic shot noise models 
that include independent, uncorrelated shots are ruled out by the observed rms-flux relations.
On the other hand, Uttley (2004) notes that such a relation can be produced by the model by \citet{L97}.  
 
 In this Paper we present the exact 
treatment of the perturbation diffusion with the generic assumption regarding the disk viscosity,  perturbation variability 
and its distribution in the disk.
 
 We should stress that in our model the emergent  signal is a result of the diffusion of the driving (local)
 perturbations in the disk  which is 
not a composition of independent, uncorrelated shots. In the  contrary, in the emergent signal all exponential shots of the composition  
are related to the spatial distribution of the driving disk perturbations and to the diffusion time scale in the disk.   Moreover, the observed rms-flux relation (Uttley 2004) is naturally produced by  the perturbation  diffusion. 
 
We also demonstrate that the specific shape of the power spectrum is determined by the distribution of the flickering
(perturbation) in the configuration, the extended disk or the CC region,  and its viscous diffusion time scale $t_0$. 
We study {\it the effect of the diffusion of these fluctuations on the accretion rate at the inner disk edge}.
Wood et al. (2001), hereafter W01, formulated and solved the problem of the  diffusive  propagation of the initial distribution of 
the perturbations (Cauchy's problem) in the  disk analytically. 
We use their general solution to derive the power spectrum of the timing signal which is a response to the diffusion of {\it the 
persistent driving} perturbations in the disk-like configuration.   

The detailed study 
of the nature of the driving perturbations of the disk is beyond the  scope of the present work.
As we mentioned above the local driving perturbations of the disk can be a result of the dynamo as a 
small-scale stochastic phenomenon,  operating on roughly the local dynamical time-scale (see K04). 
However we  suggest that they are  likely due to the Rayleigh-Taylor (RT) local instability [e.g. \citet{c61} and \citet{t03}]
 which is a common phenomenon for any configuration with non-homogeneous vertical density structure. 
These local RT instabilities  are usually seen in the Earth atmosphere as gravity waves. 

We apply the results of our theoretical investigations to the RXTE and EXOSAT observations of Cyg X-1 and 
Cyg X-2. Churazov, Gilfanov \& Revnivtsev (2001), argued using their data analysis of RXTE observations that the 
overall shape of the PDS in the soft and hard spectral states can be {\it qualitatively} explained if the geometrically thin disk 
is sandwiched by the geometrically thick CC extending in radial direction up to a large
distance from the compact object. They also suggested that in the hard state the thin disk is truncated at some distance 
from the black hole followed by the geometrically thick flow. Our quantitative analysis of PDS and photon spectra of Cyg X-1 and Cyg X-2
confirms  this  idea.  We show that the accretion flow presumably consists of a geometrically thin extended  disk component 
and a geometrically thick CC component.

GA06 analyzed the PDSs for a number of NS and BH sources using RXTE and EXOSAT observations in the wide frequency range, 
from $10^{-8}$ Hz to $10^{2}$ Hz. Particularly, they found the composite PDS of Cyg X-2. We reproduce their composite PDS. 
Furthermore, we infer the physical characteristics of the accretion flow in Cyg X-2 by application of our theory to 
the observable PDS. 

In \S 2 we present the details of timing signal formation  in the source of X-ray radiation.  
In \S 3 we detail the main features of the solution of the initial value problem which is tightly related  
to  the shape of the emergent timing signal.
We demonstrate these solutions   for specific and general cases of disk viscosity 
as a function of radius. In \S 4 we show how our theory fits X-ray data from Cyg X-1 and Cyg X-2. Specifically, we
explain how the power spectrum properties correlate with energy spectrum in different spectral states. Discussion of 
the main results and conclusions follow in \S 5 and \S 6.

\section{Evolution of the  power spectra. Theoretical consideration} 

In this Paper we study the diffusive propagation of the local dynamical perturbations (fluctuations) 
in the disk-like bounded configuration.  In other words, we assume that there is a temporal source 
of fluctuation at any point (radius) of the medium (disk) $\Phi(R, t)$ .  The X-ray time response of the disk,
 the luminosity perturbation,   $\Delta L_x(t)$
can be considered in terms of diffusive propagation of the local driving perturbations $\Phi(R, t)$ in 
the disk.  
We  assume  that the temporal local  variation of the mass supply  in the disk  
around the steady state are small.  They are only some fraction of the steady state mass supply  through the disk. 
In other words,  the amplitude of    $\Phi(R, t)$ is proportional to the steady state mass accretion rate. 
W01 show (see Eq. 7 there)  that the mass accretion rate at the innermost radius  of the disk  $\dot M(R_{in},t )$ 
is  proportional to the mass supply over the disk $A(t)$. 
It implies that  the  $\dot M(R_{in},t )$ perturbations, $\Delta\dot M(R_{in},t )$, should be proportional to 
$\Delta A(t)=2\pi \int \Phi(R, t)RdR$ and consequently proportional to $A(t)$ because 
$\Delta A(t)\propto A$.  

Thus one can formulate the problem of  the diffusive propagation of the surface density perturbations
$\Phi(R, t)$   in the bounded configuration (see Eq. \ref{dif_driv_eq} and Eq. 5 in W01).   
It is important to emphasize that this diffusive propagation of fluctuations is an intrinsic property of a given
 disk-like configuration (necessary condition) where the angular momentum is distributed by diffusion (see Eq. \ref{dif_driv_eq}).  

The resulting power spectrum as a result of the  diffusion of perturbations in the disk 
$||F_{x}(\omega)||^2$
  is a product of the power spectrum of the temporal variation of source perturbations 
  $||F_{\varphi}(\omega)||^2$ and   the power spectrum of the disk response to the spatial distribution of 
the driving perturbations  over the disk $||F_{Y}(\omega)||^2$ (see Eq. \ref{pwsp_varphi_res}).  
Because the power of the driving  perturbations $||F_{\varphi}(\omega)||^2$ 
 is  directly related to {\it the mass supply over the disk}  $A(t)$ (see above) the intrinsic property of the  
diffusive propagation  of the driving perturbation is {\it the existence of a strong  relationship between the 
amplitude of the X-ray variability,  related to $||F_{x}(\omega)||^2$,  and the X-ray flux [$\propto A(t)$]}
[compare with  the result of Uttley (2004)].
 
Thus  in the observations we see the dynamical variation in the disk as a variation of the X-ray  photon flux. 
The X-ray spectrum  is likely formed as a result of upscattering  of the soft Keplerian disk photons in the 
Compton cloud, which can   be regarded  as a disk-like sub-Keplerian configuration.  
One  also has to take into account  photon diffusion in the power spectrum formation 
(see Eq. \ref{conv}).  The emergent power spectrum  is a result of the combined effect of the CC photon diffusion and the 
diffusion of the disk fluctuations i.e.  a product of the power spectra for the photon diffusion and 
fluctuation diffusion respectively  (see Eq. \ref{pwsp}).

If  two disk-like bounded configurations are sources of the perturbation, i.e.,  $\Delta L_{x,1}(t)$ and 
$\Delta L_{x,2}(t)$ in the system 
which are weakly correlated, then the resulting power spectrum is a sum of the corresponding spectra (see Appendix  A)
\begin{equation}
||F_{x}(\omega)||\approx||F_{x,1}(\omega)||^2+||F_{x,2}(\omega)||^2.
\label{sum_pds}
\end{equation}

\subsection{Diffusive propagation of the perturbation  in the disk. Formulation of the problem}
   
Here we consider the diffusive propagation when the driving perturbations can be presented in a factorized 
form $\Phi(t,R)=\varphi(t)f(R)$. In other words   a spatial distribution of the driving perturbations in the disk is described by $f(R)$ and $\varphi(t)$ characterizes  the perturbation input rate at any disk radius.    In Appendix B.1 we show that the diffusion solution for the general case  
of the function $\Phi (t, R)$ can be well  approximated   by the solution for $\Phi(t,R)=\varphi(t)f(R)$.


The diffusion equation for the time variable quantity $W(R,t)$,  related to the surface density perturbations $\Delta \Sigma(R,t)$, $W(R,t) =\Delta\Sigma(R,t)$,  can be written in an operator form    (see Eq. 5 in W01):
\begin{equation}
{{\partial W}\over{\partial t}}={\bf\Lambda_R} W+\varphi(t)f(R)
\label{dif_driv_eq}
\end{equation}
where $R$ is a radial coordinate in the disk and ${\bf\Lambda_R}$ is the space diffusion operator.
 Equation (\ref{dif_driv_eq}) should be combined with the appropriate boundary conditions at $R=0$, $R=R_0$ and initial conditions at $t=0$. 
For homogeneous initial conditions, namely for $W(R,0)=0$ the solution at any $R$  and $t$ can be presented as a convolution 
\begin{equation}
W(R,t)=\int_0^t \varphi(t^{\prime}) X(R,t-t^{\prime})dt^{\prime}.
\label{convd}
\end{equation} 
The kernel of convolution (\ref{convd}),  $X(R,t-t^{\prime})$ is a solution of the initial value problem for the  homogeneous equation
\begin{equation}
{{\partial X}\over{\partial t}}={\bf\Lambda_R} X 
\label{hom_dif_driv_eq}
\end{equation}    
 with the following initial conditions
\begin{equation}
X(R,t-t^{\prime})_{t=t^{\prime}}=X(R,0)=f(R)
\label{init_cond0}
\end{equation}  
and with the same boundary conditions as that for $W(R,t)$ 
(we specify them in \S 3, see also Eqs. 15, 16 in W01).   
 The validity of $W(R,t)$, presented by formula (\ref{convd}), as a solution of Eq. (\ref{dif_driv_eq}) with the homogeneous initial condition can be directly checked by its  substitution to   Eq. (\ref{dif_driv_eq}) having in mind Eqs (\ref{hom_dif_driv_eq}-\ref{init_cond0}) for  $X(R,t-t^{\prime})$.

It is important to point out that resulting perturbation signal is a sum of two components where one component is 
presented by formula (\ref{convd}) and the other component is a solution of the initial value problem of a homogeneous 
diffusion equation, analogous to Eqs (\ref{hom_dif_driv_eq}-\ref{init_cond0}) but with the initial  perturbation function 
that  can be  different from $f(R)$ (see \S3).  

If  the observational  time 
intervals are   much  longer than the characteristic diffusion time scale of perturbation in the disk $t_0$ 
then the contribution of the second component  of the resulting signal is exponentially small
\footnote{Gilfanov  (2006, private communication) points out to the fast decay of this component with time.}.  
The amplitude of perturbations determined by the solution  of the homogeneous problem 
[see Eqs (\ref{hom_dif_driv_eq}-\ref{init_cond0})] decays exponentially for $t\gg t_0$ 
(see details in  \S 3). 

The power spectrum of 
$||F_{W}(\omega)||^2$ of  $W(R, t)$ 
can be presented  as a  product of the power spectra $||F_{\varphi}(\omega)||^2$ and 
$||F_X(\omega)||^2$  of $\varphi(t)$ and $X(R,t)$ respectively: 
\begin{equation}
||F_{W}(\omega,R)||^2=||F_{\varphi}(\omega)||^2||F_{X}(\omega,R)||^2,
\label{pwsp_varphi}
\end{equation}
where
\begin{equation}
F_W(\omega,R)=\frac{1}{\sqrt{2\pi}}\int_0^{\infty}e^{-i\omega t}W(t)dt,
\label{ftz}
\end{equation}

\begin{equation}
F_{\varphi}(\omega)=\frac{1}{\sqrt{2\pi}} \int_0^{\infty}e^{-i\omega t}\varphi(t)dt,
\label{ftx}
\end{equation}

\begin{equation}
F_{X}(\omega,R)=\frac{1}{\sqrt{2\pi}}\int_0^{\infty}e^{-i\omega t}X(R,t)dt
\label{ftg}
\end{equation}
are  Fourier transforms of $W(R,t), ~\varphi(t),~X(R,t)$ respectively.

The X-ray resulting variable signal is determined by  
the  fluctuations of the  luminosity $\Delta L_x(t)$.
 We assume that  
the mass accretion rate variations $\Delta \dot M(0,t)$ is converted with efficiency $\varepsilon_{eff}$ into the variations of the X-ray luminosity, i.e. $\Delta L_x(t)= \varepsilon_{eff}\Delta \dot M(0,t)$.

W01 show that  for the function ${\cal W}(x,t)=x\hat \nu W(x^2,t) $ using a new variable $x=R^{1/2}$  the diffusion equation (\ref{dif_driv_eq}) can be presented      in the form 
\begin{equation}
{{\partial {\cal W}}\over{\partial t}}=\frac{3\hat \nu(x)}{4x^2}\frac{\partial^2{\cal W}}{\partial x^2}+
\varphi(t){\cal F} (x)
\label{dif_driv_eq_mod}
\end{equation} 
where  $\hat\nu(x)$ is viscosity in the disk, ${\cal F}(x)=x\hat\nu(x)f(x^2)$.
The convolution, similar to Eq. (\ref{convd}), presents the solution ${\cal W}(x,t)$ 
 \begin{equation}
{\cal W}(x,t)=\int_0^t \varphi(t^{\prime}) {\cal X}(x,t-t^{\prime})dt^{\prime}
\label{convdm}
\end{equation}
where ${\cal X}(x,t)$ is a solution of the initial value problem 
(compare with  Eqs. \ref{hom_dif_driv_eq}, \ref{init_cond0})
\begin{equation}
{{\partial {\cal X}}\over{\partial t}}=\frac{3\hat \nu(x)}{4x^2}\frac{\partial^2{\cal X}}{\partial x^2}
\label{hom_dif_driv_eqm}
\end{equation}    
 with the following initial conditions
\begin{equation}
{\cal X}(x,0)={\cal F}(x).
\label{init_cond0mm}
\end{equation}

W01 (see Eq. 10 there) find that 
\begin{equation}
\Delta L_x(t)= \varepsilon_{eff}\Delta \dot M(0,t)=3\pi\varepsilon_{eff}\frac{\partial {\cal W}}{\partial x}(0,t). 
\label{delta_L}
\end{equation}
The   total X-ray deposition of the fluctuations  at the inner disk edge $Q_x$ can be obtained if we integrate Eq. (\ref{dif_driv_eq_mod}) over t (from $0$ to $\infty$) and over x (from $0$ to $x_0=R^{1/2}_0$). Note that the time integral in the left hand side of Eq. (\ref{dif_driv_eq_mod}) 
\begin{equation}
\int_0^{\infty} {{\partial {\cal W}}\over{\partial t}}dt= {\cal W}(x,t)|_{t=\infty}-{\cal W}(x,t)|_{t=0}=0
\label{int_Wt}
\end{equation}
because we are only interested in the solution for which ${\cal W}(x,\infty)= {\cal W}(x,0)=0$.
Thus using the integration of the right hand side of  (\ref{dif_driv_eq_mod}) combined with Eq. (\ref{delta_L}) and the outer boundary condition $\partial {\cal W}/{\partial x}(x_0,t)=0$ (see W01, Eq. 15)   
and keeping in mind the relation  $x=r^{1/2}$
we find that 
\begin{equation}
Q_x=\int_0^{\infty}\Delta L_x(t)dt =
\left [\varepsilon_{eff} \int_0^{R_0}f(R)2\pi RdR\right]\int_0^{\infty}\varphi(t)dt
=C_{dr}\int_0^{\infty}\varphi(t)dt.
\label{Q_eq}
\end{equation}
Since the  function $f(R)$  determines the shape of the spatial distribution of the driving perturbation only,  we can normalize $f(R)$ in such a way that  the factor $C_{dr}$ in the right hand side of 
Eq. (\ref{Q_eq}) is equal to 1,  namely 
\begin{equation}
C_{dr}=\varepsilon_{eff} \int_0^{R_0}f(R)2\pi RdR=1.
\label{C_S}
\end{equation}
In this case  the total  X-ray fluctuation energy  $Q_x$ is equal to the integrated input (flux) of  the driving perturbations over the disk:
  \begin{equation}
Q_x=\int_0^{\infty}\varphi(t)dt= Q_{dr}. 
\label{Q_eqm}
\end{equation}
Eq. (\ref{Q_eqm}) implies that the emergent variable flux of X-ray emission $Q_x$ is the same as the integrated input of  the driving perturbations over the disk $Q_{dr}$. In other words {\it the driving perturbation flux is conserved when the perturbations  diffuse   through the disk towards the  inner disk edge}.

On the other hand {\it the integrated power of the resulting PDS, $P_x$  is only a small fraction of the integrated power of the driving oscillations, $P_{dr}$  distributed over the disk}.  The ratio of 
$P_x/P_{dr}$ strongly depends on the diffusion timescale in the disk $t_0$ and on the characteristic frequency of the driving oscillations $\nu_{dr}$, $\nu_{dr}=\omega_{dr}/(2\pi)$.

In fact,  the resulting X-ray signal due to the diffusion of the driving perturbations is
 \begin{equation}
\Delta L_x(t)=\int_0^t \varphi(t^{\prime}) Y(t-t^{\prime})dt^{\prime},
\label{convdL}
\end{equation}
\begin{equation}
Y(t)=3\pi\varepsilon_{eff}\frac{\partial {\cal X}(0,t-t^{\prime})}{\partial x}.
\label{signal_y}
\end{equation}
To obtain Eqs. (\ref{convdL}) and  (\ref{signal_y}) we use Eqs. (\ref{convdm}) and (\ref{delta_L}).
 
Then the resulting power spectrum is 
\begin{equation}
||F_{x}(\omega)||^2=||F_{\varphi}(\omega)||^2||F_{Y}(\omega)||^2
\label{pwsp_varphi_res}
\end{equation}
where $||F_{x}(\omega)||^2,~||F_{\varphi}(\omega)||^2,~||F_{Y}(\omega)||^2$ are  Fourier transforms of $\Delta L_x(t), ~\varphi(t),~Y(t)$ respectively [see e.g. Eq. (\ref{ftz})  for definition of the Fourier transform].
The disk local  driving   oscillations    convolved with the response of the  disk-like configuration results in the emergent response of the system $\Delta L_{x}(t)$.
Ultimately,  the power spectrum $||F_{x}(\omega||^2$ of $\Delta L_{x}(t)$ carries the information on the characteristic frequencies  and the hydrodynamical structure
of the system.  

In general, the disk driving fluctuation $\varphi(t)$ can be presented as  damped quasi-periodic (driving) oscillations  for which power spectrum is Lorentzian
 \begin{equation}
||F_{\varphi}(\omega)||^2\propto [(\omega-\omega_{dr})^2 + (\Gamma_{dr}/2)^{2}]^{-1}
\label{pwsp_dampos}
\end{equation}
where  $\Gamma_{dr}$ is a damping factor.
We suggest that the frequency of the disk  driving oscillations  $\omega_{dr}$ (as a frequency of the Raylegh-Taylor gravity waves)  is scaled with  the local Keplerian frequency $\omega_{\rm K}$. 
 In fact,  $\omega_{dr}$ is some mean value of the rotational frequency of the local quasiperiodic oscillations in the disk-like configuration (see Appendix B.1).

Now we proceed with an estimate of the integrated total power of the resulting signal  $P_x=\int_0^{\infty} ||F_{x}(\omega)||^2d\omega$. Using a relation Eqs (\ref{pwsp_varphi_res}), (\ref{Y2_generalA}),
(\ref{FX_mean_prA}), (\ref{FX_meanA})
we obtain that 
 the integrated total power of the resulting signal
\begin{equation}
P_x=\int_0^{\infty} ||F_{x}(\omega)||^2d\omega\sim\frac{1}{D{\cal Q}} \frac{P_{dr}}{\omega_{dr}t_{0}}.
\label{power_L}
\end{equation}
Here ${\cal Q}=\omega_{dr}/\Delta\omega\gax 1$ stands for  a quality factor, as  $\Delta\omega$ stands for a FWHM of $||F_{\varphi}(\omega)||^2$ and a numerical factor $D\gax1$.
We emphasize  that $\omega_{dr}t_{0}\gg1$ 
because the diffusion time scale in the disk  $t_{0}$ 
is likely  much longer than the timescale of (local) driving oscillation $t_{dr}\sim \omega^{-1}_{dr}$.

Thus using   equation  (\ref{power_L}) we arrive to the conclusion that the {\it resulting integrated power $P_x$, which is  related to the perturbation amplitude  at the  inner disk edge,   is much less than
the {\it total} integrated power of the driving oscillation in the disk $P_{dr}$ }\footnote{It should be noted that the referee's question motivates us to mathematically  prove this statement (see Eq. \ref {power_L}).} 
\begin{equation}  
\frac{P_{x}}{P_{dr}}\sim(D{\cal Q}\omega_{dr}t_{0})^{-1}\ll1.
\label{PL_PX}
\end{equation}
This is a prediction of our diffusion model.  The model can be confirmed or refuted if one can determine product of $\omega_{dr}$ and $t_0$ from observations and compare this with the observed ratio of $P_x$ and $P_{dr}$.   In \S 5 we demonstrate the validity of our model using the power spectrum of Cyg X-1 obtained  in the broad frequency range, from $10^{-7}$ Hz to $10^2$ Hz.

Even though the specific mechanism providing the disk viscosity needs to be understood, the diffusion time scale in the disk and driving oscillation frequency
``controls'' 
the variability  of the innermost region of the accretion disk (Compton cloud). 
As seen from Eq. (\ref{PL_PX}) the power (rms$^2$) of the resulting disk fluctuations should decrease with $\omega_{dr}$. 
On the other  hand  it is well established (see e.g. ST06) that  the X-ray emission area (Compton cloud)  becomes more compact  when the X-ray source evolves from hard to soft states. Average driving oscillation frequency $ \omega_{dr}$ should progress to higher values  during this evolution because  
 it is scaled with some mean  $\omega_{\rm K}$ over  the Compton cloud.   Probably  this decrease 
 of $P_{x}/P_{dr}$ with $\omega_{dr}$ is a key to  explain   the rms decay when the X-ray source evolves from hard to soft states (and  when $\omega_{dr}$    increases).




\subsection{Effect of the photon diffusion in Compton Cloud on the emergent power spectrum}
 
ST06  revealed that the PDS features,  
break frequency $\nu_b$, and Lorentzian low-frequency $\nu_L$ and Q-value of the QPO frequency evolve and increase while the 
source progresses toward the high-soft state. But  the QPO frequencies  are completely washed out in the  very soft state. 
Titarchuk et al. (2002), hereafter TCW02, predicted that when the source is embedded in the optically thick   
medium the QPO features must be absent in the PDS of the source because of photon scattering. 




The strength of QPO frequencies can be easily  attenuated when the oscillating X-ray emission passes through the intervening   wind environment.  The power of the wind presumably related to the mass accretion rate in the disk.  In Appendix  E we present the details of derivation of  the wind outflow rate and optical depth.  We also show there that the mass outflow rate is much higher than the mass accretion rate in sources, like Cyg X-1, Cyg X-2 and Sco X-1, with the high supply of the matter from the companion.

Thus we can conclude that the variability of the scattered part of radiation can be  completely washed out  in the extended wind, of radius of order $10^{11}$ cm for Cyg X-1,  
even for the wind optical depth $\tau_W\gax0.5$ (ST06). 
The variability of the direct (unscattered) component is preserved but its rms amplitude  
decreases as $\exp(-\tau_W)$  with $\tau_W$. ST06's  data analysis confirms this expectation. 
The power spectrum in the very soft state 
is featureless (see ST06, Fig. 6). The emission of the central source is presumably obscured by 
the optically thick wind and consequently 
all photons emanating from the central source are scattered. The direct component of the central source 
radiation that carries information about the variability is suppressed by scattering. 

\citet{gies} argue that there is a particular state of Cyg X-1 when the wind velocity is very low, and 
thus one can expect high accumulation material in the wind and noticeable optical depth of the wind (see also Eq. \ref{tau outflow2}).  The wind 
downscattering of the photons emanating from the inner Compton cloud leads to the softening of spectrum \citep{tsh05} 
and consequently to decrease of X-ray luminosity of the source. The softening of the spectrum can also be 
a result of effective cooling the Compton cloud by the disk soft photons.
If the power law spectra in the soft state are formed in the converging flow, then their indices are determined by 
the flow temperature (Laurent \& Titarchuk 1999, hereafter LT99). 
The index  increases and  saturates to the critical value about 2.8 with the mass accretion rate  for the low 
temperatures of the flow (see LT99 and Titarchuk \& Zannias 1998).  

We can calculate the emergent power spectrum as a result of the diffusive propagation of the perturbations in the 
disk-like configuration (Keplerian disk or sub-Keplerian Compton Cloud) and   the diffusive X-ray photon propagation from  
the innermost part of the source.  
The quantitative model of the resulting pulse affected by these perturbations  $Z(t)$  can be written as follows:
\begin{equation}
Z(R,t)=\int_0^t \Delta L_x( t^{\prime}) G(t-t^{\prime})dt^{\prime}
\label{conv}
\end{equation}
where $\Delta L_x(t)$ is the input pulse of the disk-like source oscillations and  $G(t)$ is the photon response pulse of the 
transition layer (TL). Then the observed power spectrum of $Z(t)$ is a product of the power spectrum of $\Delta L_x(t)$ 
(see Eq. \ref{sum_pds})
and $G(t)$, i.e.
\begin{equation}
||F_Z(\omega)||^2=||F_{x}(\omega)||^2||F_{G}(\omega)||^2.
\label{pwsp}
\end{equation}



\section{Initial value  (Cauchy) problem of diffusion.  Diffusion of  the radial local  perturbations in the disk  as an origin of the ``white-red'' noise. The analytical solution of the problem: General and Particular cases}

The diffusive propagation of the perturbation 
in the disk was studied by W01 in detail.  They presented  the diffusion equation for the surface density 
$\Sigma (R,t)$  as a function  of time $t$ and the radial position in the disk $R$ (see Eq. 5 in W01). It is worth noting that 
 the derived equation is valid for any  disk-like configuration for which the rotational frequency profile is Keplerian.  
This configuration can be 
a Shakura-Sunyaev type of  disk (Shakura \& Sunyaev 1973) or  an advection dominated accretion flow (ADAF) (Narayan \& Yi 1994, 
Chakrabarti \& Titarchuk 1995). Thus this applies to the Compton cloud as well.  

It is also important to emphasize that  the  equation derived for the surface density  can be 
used as a   equation for the surface density perturbations $\Delta\Sigma(R,t)$   in framework of the  linear perturbation theory 
(see  also \S 2.1).
In section \S 2 {\it we demonstrate that the determination of the diffusion response of the disk to the driving oscillations 
$\varphi(t)f(R)$ is reduced  to the convolution of $\varphi(t)$ with 
 the solution of the initial value (Cauchy) problem for the  distributed  perturbations at the initial moment $f(R)$} 
[see Eqs. (\ref{hom_dif_driv_eq}, \ref{init_cond0})].  
The initial value (Cauchy) problem of the time-dependent diffusive propagation of the surface density perturbation is  
formulated as follows (see also Eqs. \ref{hom_dif_driv_eqm}, \ref{init_cond0mm}):
\begin{equation}
{{\partial {\cal X}}\over{\partial t}}={\bf\Lambda_x} {\cal X}={{3{\hat\nu}(x)}\over{4x^2}} 
{{\partial^2{\cal X}}\over {\partial x^2}}
\label{main_eq}
\end{equation}
with the initial condition 
\begin{equation}
{\cal X}(x)={\cal F}(x)=x\hat\nu(x)f(x^2)~~~~{\rm at}~~~~t=0
\label{init_cond}
\end{equation}
where $x=R^{1/2}$,  $\hat\nu(x)$ is a  viscosity in the disk and 
\begin{equation}
{\cal X}(x,t)=R^{1/2}\hat\nu W=R^{1/2}\hat\nu\Delta\Sigma.
\label{v_func}
\end{equation}

We  combine  equation (\ref{main_eq}) with the boundary condition at the outer boundary
\begin{equation}
{{\partial {\cal X}}\over {\partial x}}=0~~~~{\rm at}~~~x=x_0
\label{boun_out}
\end{equation}
and  at the inner boundary $x_{\rm in}\ll x_0$,
$W=\Delta \Sigma=0$, which is equivalent to
\begin{equation}
{\cal X}=0~~~~~{\rm at}~~~~ x=x_{\rm in}.
\label{boun_in}
\end{equation}
W01 demonstrated how  the mass accretion rate in the disk $\dot M$ can be calculated   
(see Eq.10  in W01). 
Using  the W01's formula and the  definition of ${\cal X}(x,t)$ in Eq. (\ref{v_func}) we can determine  of the mass accretion rate perturbation as follows
\begin{equation}
\Delta \dot M= 3\pi 
{{\partial {\cal X}}\over {\partial x}}.
\end{equation}
Furthermore, we assume that the perturbation of the mass accretion rate  at the inner disk
edge is converted with efficiency $\varepsilon_{\rm eff}$ into the perturbation of 
X-ray luminosity, $\Delta_xL(t)$ i.e.  $\Delta L(t)= \varepsilon_{\rm eff}\Delta \dot M(t,R_{in})$ and thus 
\begin{equation}
Y(t)\propto\Delta L_x(t)\propto {{\partial {\cal X}}\over {\partial x}}(t,0).
\label{Y_signal}
\end{equation}
The solution ${\cal X}(x,t)$ of equation (\ref{main_eq}) with the initial condition (\ref{init_cond}) at
$t=0$ and boundary conditions (\ref{boun_out}-\ref{boun_in}) can be presented using 
separation of variables as a series
\begin{equation}
{\cal X}(x,t)=\displaystyle\sum_{k=1}^{\infty} 
e^{-\lambda_k^2 t}{{\chi_k(x)c_k}\over{||\chi_k(x)||^2}},
\label{V_series}
\end{equation}
where $\chi_k(x)$ and $\lambda_k$ are eigenfunctions and eigenvalues which
can be found from the homogeneous ordinary differential
equation:
\begin{equation}
\chi_k^{\prime\prime}+\lambda_k^2p(x)\chi_n=0
\label{eigen_diffeq}
\end{equation}
combined with the boundary conditions
\begin{equation}
\chi_k=0~~~~ {\rm for}~~~~x\rightarrow 0,
\label{eigen_in}
\end{equation}
\begin{equation}
{{d \chi_k}\over{d x}}=0~~~~{\rm at}~~~~x=x_0.
\label{eigen_out}
\end{equation}
$||\chi_k||$ is the norm of the eigenfunction, which is calculated through the 
integral [for example, see the derivation of this formula in Titarchuk, Mastichiadis \&  Kylafis (1997)]
\begin{equation}
||\chi_k||^2=\int_0^{x_0}p(x)\chi_k^2(x)dx
\end{equation}
where $p(x)=4x^2/3\hat\nu(x)$ is the weight function and the expansion coefficient 
\begin{equation}
c_k=\int_{0}^{x_{0}}p(x)\chi_k(x){\cal F}(x)dx.
\label{expcoef}
\end{equation} 

Let us express $c_k$ using eigen-value equation (\ref{eigen_diffeq})
\begin{equation}
c_k=\int_{0}^{x_{0}}p(x)\chi_k(x){\cal F}(x)dx=-\frac{1}{\lambda_k^2}\int_{0}^{x_{0}}\chi_k^{\prime\prime}(x){\cal F}(x)dx.
\label{expcoef1}
\end{equation} 
By integration by parts and using boundary conditions (\ref{eigen_in}-\ref{eigen_out}) for $\chi_k(x)$ we obtain
\begin{equation}
c_k=\frac{1}{\lambda_k^2}
[\chi_k^{\prime}(0){\cal F}(0)+\chi_k(x_0){\cal F}^{\prime}(x_0)-\int_{0}^{x_{0}}\chi_k(x){\cal F}^{\prime\prime}(x)dx].
\label{expcoef2}
\end{equation}
It is evident from Eq. (\ref{expcoef2}) that 
\begin{equation}
c_k\sim\frac{1}{\lambda_k^2}\chi_k^{\prime}(0){\cal F}(0),
\label{expcoef3}
\end{equation}
for a quasi-iniform initial distribution ${\cal F}(x)$, i.e. for which 
its values and  derivatives are bounded. In fact $\chi_k^{\prime}(0)\propto \lambda_k^{\zeta}$, where $\zeta>0$ (see a proof of this
below), and  $\lambda_k=O(k)$ with increase of $k$ and thus the first term of the right hand side in 
Eq. (\ref{expcoef2}) is dominant for $k\gg1$.
It is worth noting that formula  (\ref{expcoef3}) is exact for ${\cal F}(x)={\cal F}(0)=constant$, namely for the uniform initial distribution of perturbations. 
 
 Now we study  a case of problems where $\hat\nu(x)=(\hat\nu_0/x_0^{\psi})x^{\psi}$.
\subsection{Case with $\psi=2$}

W01 showed that in this case
\begin{equation}
\chi_k(x) =\sin\left[{2\over{(3\hat\nu_0/x_0^2)^{1/2}}}\lambda_{k}x\right],
\label{eigenf_psi=2}
\end{equation}
and 
\begin{equation}
\lambda_k={{(3\hat\nu_0/x_0^2)^{1/2}}\over 2}{{\pi(2k-1)}\over{2x_0}}.
\label{eigenv_psi=2}
\end{equation}
For this particular case the time dependent signal $Y(t)$ is presented as
[see Eqs. (\ref {Y_signal}) and (\ref{V_series})]
\begin{equation}
Y(t)\propto\sum_{k=1}^{\infty}
[\pi(2k-1)/2](c_k/||\chi_k||^2)\exp[-{{\pi^2(2k-1)^2t}/4t_0}]
\label{y_signal0}
\end{equation}
where $t_0=4x_0^4/3\hat\nu_0=4R_0^2/3\hat\nu(R_0)$.  $t_0$ is the viscous
timescale and determines both the rise and fall time of the response function $Y(t)$ (see details in W01).

For the uniform perturbation source distribution $({\cal F}(x)=constant)$ (see Eq. \ref{y_signal0}) we obtain that 
\begin{equation}
Y(t)\propto\sum_{k=1}^{\infty}\exp[-{{\pi^2(2k-1)^2t}/4t_0}].
\label{y_signal1}
\end{equation}
because
\begin{equation}
||\chi_k||^2=\frac{4x_0^2}{3\hat\nu_0}
\displaystyle\int_0^{x_0}\sin^2\left[{{\pi(2k-1)x}\over{2x_0}}\right]dx=\frac{2x_0^3}{3\hat\nu_0}
\label{norm_2}
\end{equation}
and 
\begin{equation}
c_k=\displaystyle\frac{2x_0}{(3\hat\nu_0)^{1/2}\lambda_k}{\cal F}(0).
\label{expcoef_uniform}
\end{equation}
Using  relation (\ref{sc_appr_power}) for the power spectrum 
$||F_Y(\omega)||^2$  of $Y(t)$ we obtain that 
\begin{equation}
||F_Y(\nu)||^2_{\nu}\propto\sum_{k=0}^{\infty}\frac{1}{(8t_0\nu/\pi)^2+(2k+1)^4}. 
\label{power_spectrumy}
\end{equation}
It is worth noting that in Eq. (\ref{power_spectrumy})  frequencies $\nu$ are scaled with the diffusion frequency $1/t_0$.  
The result of  summation in of Eq. (\ref{power_spectrumy})   can be presented by an  exact analytical formula [see  Prudnikov, 
Bruchkov \& Marichev (1981), hereafter PBM81,  formulas (5.1.28.1)]. Thus 
$$
||F_Y(\nu)||^2_{\nu} \propto \frac{\pi}{2^{3/2}{a}^{3/2}}\frac{\sinh2^{1/2}\pi{a}^{1/2} +\sin2^{1/2}\pi{a}^{1/2}}
{\cosh2^{1/2}\pi{a}^{1/2}- \cos2^{1/2}\pi{a}^{1/2}} -
$$
\begin{equation}
-\frac{\pi}{2^{5/2}{a}^{3/2}}\frac{\sinh\pi{a}^{1/2}/2^{1/2} +\sin\pi{a}^{1/2}/2^{1/2}}
{\cosh\pi{a}^{1/2}/2^{1/2} -\cos\pi{a}^{1/2}/2^{1/2}}
\label{power_spectrum_y1}
\end{equation} 
where $a=8t_0\nu/\pi$.
As it follows from this formula that 
\begin{equation}
||F_Y(\nu)||^2_{\nu}=C_N\times \pi^4/96~~~~~ {\rm when}~~~ \nu\ll\pi/8t_0
\label{pw_sp_lowfr}
\end{equation}
and 
\begin{equation}
||F_Y(\nu)||^2_{\nu}=C_N\times \frac{1}{2^7\pi^{1/2}t_0^{3/2}}\frac{1}{\nu^{3/2}}~~~~~ {\rm when}~~~ \nu\gg\pi/8t_0.
\label{pw_sp_highfr}
\end{equation}

If the source distribution ${\cal F}(x)$ is non-uniform, i.e. ${\cal F}(x)\neq constant$ then in the right hand side of formula (\ref{power_spectrumy}) 
an additional term appears (see Eqs. \ref{expcoef2}-\ref{expcoef3} for $c_k$) that is  order of $O(1/\nu^2)$ when $\nu\gg\pi/8t_0$. It is a constant for $\nu\ll\pi/8t_0$.

Thus if $|{\cal F}(x|, ~|{\cal F}^{\prime}(x)|, ~|{\cal F}^{\prime\prime}(x)|$ are the same order of magnitude then the sum shown in formula (\ref{power_spectrumy})
is still the leading term in the corresponding formula for the power spectrum for the non-uniform source distribution. Consequently, the index of the power-law part
of $||F_Y(\nu)||^2_{\nu}$ is still equal to $3/2$.

\subsection{General Case}
In this section we investigate the  behavior of the power
spectrum  in the case of arbitrary disk viscosity as a function of $R$ (or $x = \sqrt{R}$).  We have 
already demonstrated for the $\psi = 2$ case that the power spectrum is a constant (the white noise) at very low frequencies ($\nu\ll\pi/8t_0$) and the power law
with index $3/2$ at high frequencies $\nu\gg\pi/8t_0$. This shape of the power spectrum is practically independent of the quasi-uniform initial  distribution ${\cal F}(x)$.
We showed  that the power spectrum for the $\psi = 2$ case is presented as a series.  The calculation of the series is reduced to 
analytical formula (\ref{power_spectrum_y1}) from which low and high-frequency asymptotes are evident (see formulas \ref{pw_sp_lowfr}, \ref{pw_sp_highfr}).
The similar presentation and asymptotic form of the power spectrum  can be obtained in a general case of the
disk viscosity.

W01 showed that (see Eq. 38 there)
\begin{equation}
||\chi_k(x)||^2\simeq \mu(x_0)/2.
\end{equation}
where $\mu(x)=[2/(4-\psi)](4x_0^{\psi}/3\hat\nu_0)^{1/2}x^{(4-\psi)/2}$.
It implies that the normalization of the eigen functions $||\chi_k(x)||^2$ does not depend on $k$.
On the other hand for small values of the argument $x$, the
eigenfunction $\chi_k$ can be presented as,
\begin{equation}
\chi_k(x)=\left[{{\pi\lambda_k}\over{(4-\psi)}}\right]^{1/2}
x^{1/2}\left[\mu(x)\lambda_k/2\right]^{1/(4-\psi)}/\Gamma[(5-\psi)/(4-\psi)].
\label{eigen_small}
\end{equation}
In fact, it follows from this equation that 
\begin{equation}
\chi_k(x) \propto \displaystyle\lambda_k^{(6-\psi)/2(4-\psi)}x
\end{equation}
when $\lambda_n\mu(x)\ll 1$.
Thus 
\begin{equation}
[\chi_k^{\prime}(0)]^2 \propto \displaystyle\lambda_k^{(6-\psi)/(4-\psi)}.
\end{equation}
The square of $\chi_k^{\prime}(0)$ is used for calculation of the  signal  $Y(t)$ (see formulas  \ref{Y_signal}, \ref{V_series}, \ref{expcoef2}).

Consequently for the uniform perturbation source distribution ${\cal F}(x)=constant$  
(see Eq. \ref{expcoef3}) we obtain that $Y(t)$ can be written as a series (cf. Eq. \ref{y_signal0}) 
\begin{equation}
Y(t) \propto {{\partial {\cal X}}\over{\partial x}}(0,t)\propto
\sum_{k=1}^{\infty}e^{-\lambda_k^2t}
\lambda_k^{\delta}
\label{Y_signal_general}
\end{equation} 
where 
\begin{equation}
\delta=(\psi-2)/(4-\psi)
\label{delta} 
\end{equation}
($\delta=0$ for $\psi=2$) and 
\begin{equation}
\lambda_k = \pi[2k-(10-3\psi)/2(4-\psi)-\varepsilon_k/\pi]/2t_0^{1/2},
\label{eigen_value_gen}
\end{equation}
 
\begin{equation}
\varepsilon_k={{(2-\psi)}
\over{(4-\psi)[(k-1)\pi+\pi(6-\psi)/4(4-\psi)]}}.
\label{epsilon_k}
\end{equation}
As in the case with $\psi = 2$,  $t_0$  is the viscous time which in the 
general case is
\begin{equation}
t_0=\mu^{2}(x_0)={4\over{3\hat\nu(R_0)}}{4\over{(4-\psi)^2}}R_0^2.
\label{t0}
\end{equation} 
One can use  equations (\ref{eigen_value_gen}, \ref{epsilon_k})  for calculations of the eigenvalues and ultimately for calculation
of the power spectra series (cf. Eq. \ref{power_spectrumy}): 
\begin{equation}
||F_Y(\nu)||^2_{\nu}\propto\sum_{k=1}^{\infty}\frac{(2t_0^{1/2}\lambda_k/\pi)^{\delta}}{(8t_0\nu/\pi)^2+ (2t_0^{1/2}\lambda_k/\pi)^4}
\label{power_spectrumy_general}
\end{equation}
where
\begin{equation}
2t_0^{1/2}\lambda_k/\pi=2k-(10-3\psi)/2(4-\psi)-\varepsilon_k/\pi
\label{2k-1_mod}
\end{equation}
which equals $(2k-1)$ for $\psi=2$ (see Eq. \ref{power_spectrumy}).
Although the series of power spectrum
\begin{equation}
||F_Y(\nu)||^2_{\nu}\propto\sum_{k=1}^{\infty}\frac{[2k-(10-3\psi)/2(4-\psi)-\varepsilon_k/\pi]^{\delta}}{(8t_0\nu/\pi)^2+ [2k-(10-3\psi)/2(4-\psi)-\varepsilon_k/\pi]^4}
\label{k_series}
\end{equation}
has to be calculated numerically the asymptotic forms of $||F_Y(\nu)||^2_{\nu}$ can be easily evaluated analytically:
\begin{equation}
||F_Y(\nu)||^2_{\nu}=C_N\times {\cal A}_L~~~~~ {\rm when}~~~ \nu\ll\pi/8t_0
\label{g_pw_sp_lowfr}
\end{equation}
and 
\begin{equation}
||F_Y(\nu)||^2_{\nu}=C_N\times \frac{{\cal A}_H}{\nu^{(3-\delta)/2}}~~~~~ {\rm when}~~~ \nu\gg\pi/8t_0
\label{g_pw_sp_highfr}
\end{equation}
where
\begin{equation}
{\cal A}_L=\sum_{k=1}^{\infty}\frac{1}{[2k-(10-3\psi)/2(4-\psi)-\varepsilon_k/\pi]^{4-\delta}}
\label{A_L}
\end{equation}
and
\begin{equation}
{\cal A}_H=\frac{1}{2(8t_0\pi)^{(3-\delta)/2}}\int_0^{\infty}\frac{x^{\delta}dx}{1+x^4}.
\label{A_H}
\end{equation}

Similarly to the $\psi=2$ case the series (\ref{k_series}) is the leading term in 
the general case power spectrum for the non-uniform perturbation source distribution ${\cal F}(x)$.
For a given  $\psi$ the index of the power-law part
of the power spectrum (see Eqs. \ref{delta}, \ref{g_pw_sp_highfr})
\begin{equation}
\alpha=(3-\delta)/2=(7-2\psi)/(4-\psi). 
\label{index_pds}
\end{equation}
As we have already emphasized  in \S3.1 that  the power spectral density $||F_Y(\nu)||^2_{\nu}$ is a function of 
dimensionless  frequency of $\nu t_0$ only (see Eq. \ref{power_spectrumy} and  Eqs. \ref{k_series}-\ref{A_H} ). 

\begin{figure}[ptbptbptb]
\includegraphics[width=5in,height=7.1in,angle=-90]{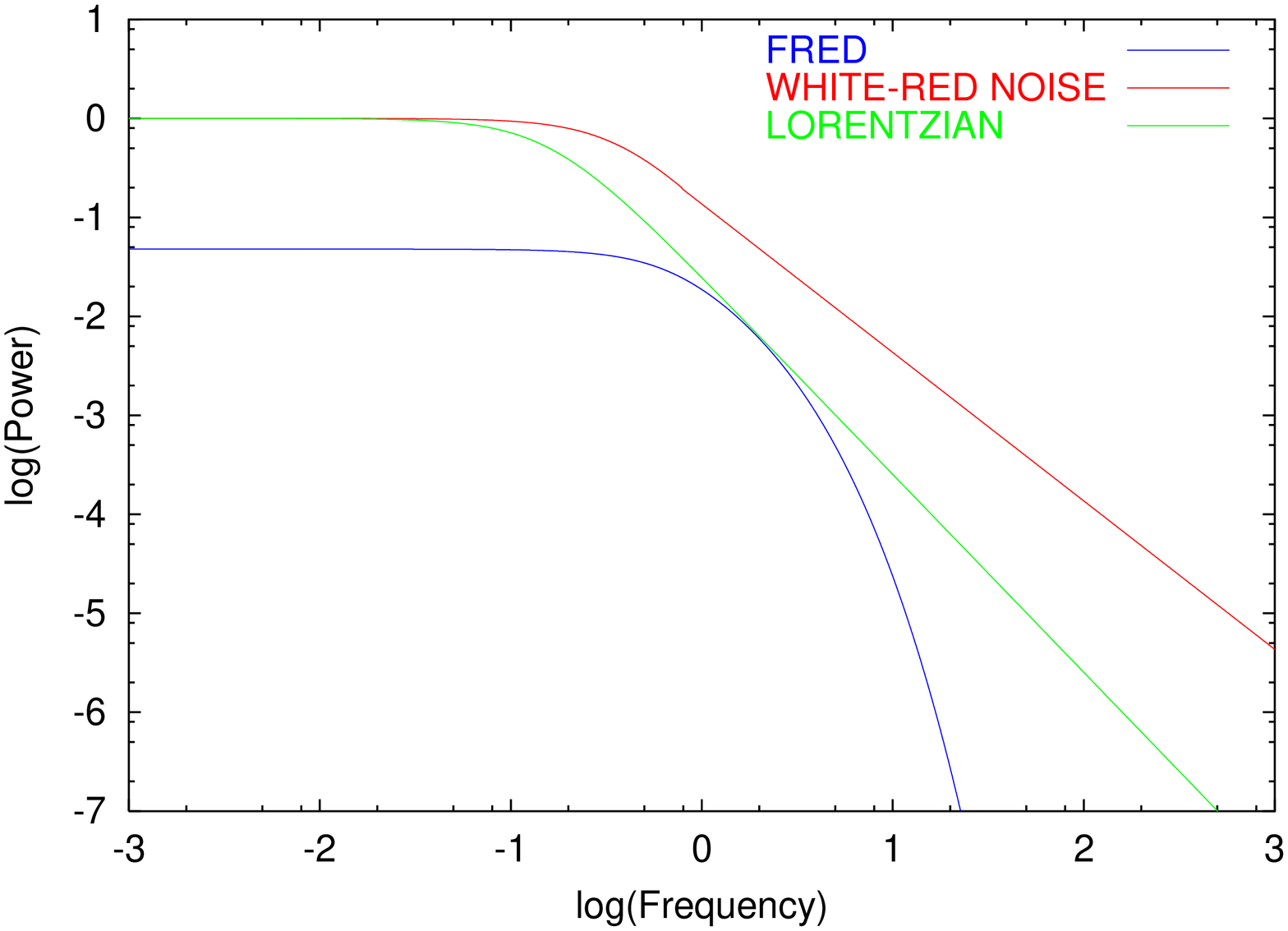}
\caption{Examples of PDS models:  PDS of fast rise and exponential decay (FRED) (blue line),
PDS of white-red noise (red line) and Lorentzian PDS (green line). 
 }
\label{pds_models}
\end{figure}

In Figure \ref{pds_models} we show the example of white-red noise (WRN) PDS calculated using formula (\ref{k_series}) for $\psi=2$. 
One can clearly see the low-frequency asymptotic form (white-noise shoulder) and high-frequency asymptotic form (red-noise power law with index 3/2,
see Eqs \ref{g_pw_sp_lowfr}-\ref{g_pw_sp_highfr})  there. For comparison we also show the PDS of an exponential shot which has a Lorentzian shape.
For calculation of the WRN PDS we also  use analytical formula (\ref{power_spectrum_y1}), that is valid when $\psi=2$,
 to verify an accuracy of the series summation using formula  (\ref{k_series}). 

\subsection{Power spectrum of the signal of the FRED type}
If the source of the perturbation is originated in the outer boundary of a given configuration
then as shown by 
W01  the response function $Y(t)$ is characterized by the fast rise and exponential decay function, namely by FRED type burst function (see Eq. 62 in W01):
\begin{equation}
Y(t)\propto (C_0+C_1/t^{\gamma+1/2})\exp(-t_0/4t-z_1^2t/t_0)
\label{FRED_gt}
\end{equation}
where 
\begin{equation}
C_0=\sin(\pi\gamma/2+\pi/4)+(z_1^2)^{\gamma/2}\cos[2(1-\gamma)/\pi(\gamma-3)],
\label{C_0_coef}
\end{equation}
\begin{equation}
C_1= 2.5\pi^{-1/2}(t_0/2)^{\gamma+1/2},
\label{C_1_coef}
\end{equation}
\begin{equation}
\gamma=(6-\psi)/2(4-\psi),
\label{gamma_coef}
\end{equation} 
\begin{equation}
z_1^2=4u_1/(4-\psi),
\label{z_1}
\end{equation}
and 
\begin{equation}
u_1=(5-\psi)\{1-[1-2/(5-\psi)]^{1/2}\}.
\label{u_1}
\end{equation}

The power spectrum 
\begin{equation}
||F_Y(\omega)||^2=C_N\{C_0^2|I_0|^2+C_1^2|I_1|^2+C_0C_1[I_0\overline{I_1} +I_1\overline{I_0}]\}
\label{gt_power_king}
\end{equation}
where
\begin{equation}
|I_j|^2=I_j\overline{I_j}=(\pi/2)\frac{2^{2(1-\alpha_j)}t_0^{2\alpha_j}}{\rho^{\alpha_j+1/2}}\exp[-2\rho^{1/2}\cos(\varphi/2)],~~~{\rm for}~~j=0,~1,
\label{I_j_modul}
\end{equation}

\begin{equation}
I_0\overline{I_1} +I_1\overline{I_0}=(\pi/2)\frac{2^{(3-\alpha_0-\alpha_1)}t_0^{\alpha_0+\alpha_1}}{\rho^{(\alpha_0+\alpha_1)/2+1/2}}
\cos[(\alpha_1-\alpha_0)\varphi/2]
\exp[-2\rho^{1/2}\cos(\varphi/2)],
\end{equation}
\begin{equation}
\rho=(z_1^4+ \omega^2 t_0^2)^{1/2},
\label{modul_rho}
\end{equation}
\begin{equation}
\varphi=\arcsin(\omega t_0/\rho)
\end{equation}
and $\alpha_0=1$ and $\alpha_1=1/2-\gamma$. 
The derivation of formula for $I_j$  is presented in the Appendix D (Eq. \ref{I_j_integral}). 
In Figure \ref{pds_models} we show the example of the FRED PDS. One can see that the FRED PDS exponentially decreases with  frequency and  this decay  is  much faster than that for
WRN and Lorentzian PDSs.

\subsection{The photon (perturbation) diffusive  propagation}
 Any local (photon or hydrodynamical) perturbation in the bounded medium would propagate diffusively outward over time scale 
 (see,  Sunyaev \& Titarchuk
 1980, hereafter ST80 and previous sections for photon and perturbation propagation respectively)
\begin{equation}
t_{\ast}\sim f\frac{l_{fp}}{v}\left(\frac{L}{l_{fp}}\right)^2=f\tau_{pert}\frac{L}{v},
\label{cstime}
\end{equation}
where $L=R_{out}-R_{in}$ is the characteristic thickness of the (photon or hydrodynamical) diffusion configuration,  
$l_{fp}\sim \eta/(\rho v)=(\sigma_{pert} n)^{-1}$ is the mean free perturbation path, related to  
 the number density $n$, the interaction cross-section $\sigma_{pert}$ in the medium, $\tau_{pert}=L/l_{fp}=\sigma_{pert} n L$ and $f$ is a factor
 which is less than 1 and its exact value determined by the space distribution of photons (perturbations) in the medium (ST80, Sunyaev \& Titarchuk 1985,
 hereafter ST85). 
 
 For the diffusive propagation in the bounded medium the response can be also presented 
 as a linear combination of the exponents (see Eq. \ref{V_series}, ST80, ST85)   
\begin{equation}
G(t)= \sum_{k=1}^{\infty}A_ke^{-\lambda_k^2t}
\label{xtdiff}
\end{equation}
where $\lambda_k$ are the eigen values related to the eigen functions 
$\chi_{k}(R)$ of the appropriate space diffusion operator, $A_k=c_k\chi_k(R_{in})/||\chi_k||^2$,
 $c_k/||\chi_k||^2$ is an expansion coefficient of the initial photon (perturbation) function $f(R)$.  As we have already shown (see e.g.  Eqs. \ref{eigenv_psi=2}-\ref{y_signal0})
$\lambda_k^2=a_k^2/t_{0}$.



 For the smooth perturbation source distribution $f(R)$ [all derivatives of $f(R)$  are bounded] 
 and   for  $\tau_{pert}$  of order of one,  
the response function as a solution of the diffusion problem can be presented by a single exponent (see ST80), namely 
\begin{equation}
G(t)\approx A_1e^{-a_1^2t/t_{0}}=A_1e^{-t/t_{\ast}}
\label{gtdiff0}
\end{equation}
because   $\lambda_1^2\ll\lambda_k^2$, for $k=2,3,...$, where $t_{\ast}=t_{0}/a_1^2$ (for example  $a_1^2=\pi^2/4$ for $\psi=2$). 

It is important to emphasize that the response of the diffusive propagation to the initial source distribution 
of  photons (or local hydrodynamical perturbations) $G(t)$ is exactly  
$A_1e^{-t/t_{\ast}}$ if  the source function $f(R)$ is proportional  to the first eigenfunction of the space diffusion operator $\chi_{k}(R)$, i.e. $f(R)\propto \chi_{k}(R)$.


For this particular response function (see Eqs. \ref{gtdiff0}) the power spectra is Lorentzian
\begin{equation}
||F_{G}(\nu)||^2_{\nu}\propto [(\nu^2 + (2\pi t_{\ast})^{-2}]^{-1}.
\label{TLpower0}
\end{equation}

In Figure \ref{pds_models}, we present the Lorentzian PDS along with the WRN and FRED PDSs. It is worth noting that the Lorentzian power-law index is fixed at 2,
as  for others the PDS power-law index is a function of the viscosity power-law index $\psi$.
 One should take into account  the effect of  photon diffusive propagation in the resulting PDS if the photon diffusion time scale is 
 comparable with the
hydrodynamical time scale in the medium. This effect is particularly important for the analysis and interpretation of the high frequency component of 
 the  PDS.


\bigskip
\centerline{\it Resulting power spectrum}
When both perturbation diffusion and photon propagation are taken into account the resulting power spectrum can be presented 
as a product (see Eq. \ref{pwsp})
\begin{equation}
||F(\omega)||^2=||F_{x}(\omega)||^2 ||F_{G}(\omega)||^2,
\label{emerg_PDS_conv}
\end{equation}
where the power spectrum of the hydrodynamical response of the source is a sum of 
\begin{equation}
||F_{x}(\omega)||^2=||F_{Y_{in}}(\omega)||^2||F_{\varphi_{in}}(\omega)||^2+
||F_{Y_{out}}(\omega)||^2 ||F_{\varphi_{out}}(\omega)||^2
\label{emerg_PDS_sum}
\end{equation}
if the density fluctuations of the inner  disk-like component (TL, Compton cloud) 
are weakly correlated 
with the fluctuations  in the extended Keplerian disk 
(see Eqs. \ref{sum_pds}, 
\ref{pwsp_varphi_res}) 

. 

\section{Applications of the theory  to Cyg X-1 and Cyg X-2 data}

We now apply the diffusion  models to a data sample from the well-studied black hole X-ray binary  Cyg X-1. The sample includes 
observations for all spectral states, 
from low-hard to high-soft states. An identification number of each RXTE observation of this sample is presented in Table 1. 
We also show a data sample for Cyg X-2 which is a neutron star (NS) source. Cyg X-2 was always in the high/soft state 
during RXTE and EXOSAT observations.

\subsection{Observations}

For our analysis we used Cyg X-1 and Cyg X-2  data from the Proportional Counter Array (PCA) and
All-Sky  Monitor (ASM) onboard {\it RXTE} (Swank, 1999) and  the medium energy (ME) detector of EXOSAT satellite 
(Turner, Smith \& Zimmermann 1981). The data are available through 
the GSFC public archive \footnote{http://heasarc.gsfc.nasa.gov}.  A reader can find the details of Cyg X-1 
observations during the entire {\it RXTE}  era in ST06. 
These data cover the period 1996 - 2006 (MJD range $\sim$ 50100 - 53800). 

The ASM instrument operates in the 2 - 12 keV energy
range at 3 energy channels and performs
flux measurements once per satellite orbit, i.e. every $\sim 90$ min. Each flux
measurement (dwell) has duration of $\sim90$ s. Due to
navigational constraints and appearance of very bright transient sources, the
ASM light curve for particular source
sometimes has gaps of duration  up to a few months. The dwell-by-dwell
light curves at 3  energy channels have been retrieved from the public RXTE/ASM archive at HEASARC. 
EXOSAT provided up to several tens of ks long light curves with a typical time
resolution of $\sim1$ s in the $0.9-8.9$ keV energy range.
The EXOSAT data for long-lasting observations  have been also retrieved from
HEASARC. Their PDSs were computed in the $2-12$ keV (ASM) and $0.9 - 8.9$ keV
(EXOSAT ME) energy range. The
PDSs of the sources from ASM lightcurves were obtained using the method based
on the autocorrelation function calculation  described by GA06. The EXOSAT light curves
were analyzed with the powspec task from FTOOLS 5.2. 
In analyzing the EXOSAT data for that individual PDSs which were similar in
shape and normalization, we averaged  them  to achieve better
statistics.

The noise level, calculated for ASM power spectra, although approximately
correct, is not accurate enough,
due to existence of unaccounted systematic errors in the flux measurements
(e.g. Grimm et al. 2002, GA06).
This leads to overestimation of values for high frequency part of  PDS.
This high frequency overestimation can also be related
to the specifics of ASM light curves, namely, it can be caused by the
aliasing effect leading to the leakage of higher frequency power below the
Nyquist frequency. Because for EXOSAT light curves the noise level is not
an issue  and normalization of EXOSAT PDS spectra are correct we have
estimated the ASM noise level using EXOSAT values of PDS for overlapping (for
both missions) frequency range. This procedure is reliable for Cyg X-2
data for which PDSs, found from individual EXOSAT sessions,   are very similar.

For Cyg X-1 PDSs of  individual EXOSAT observation are quite
different in shape and amplitude, therefore one should be more careful to smoothly
connect ASM and EXOSAT data.  
  
  It is known that Cyg X-1 often goes to state transition from regular
low/hard state to rarer soft state and vice versa (see ST06 for more details of Cyg X-1 spectral state  history).  To avoid   the influence
of such transitions on the composite EXOSAT-ASM power spectrum we have separately calculated PDS for
low/hard and soft states of Cyg X-1.  To identify  a spectral  state we
have calculated the power-law index of  the photon spectrum $ \Gamma$ based on ASM
data from different energy channels (Smith et al. 2002). For low/hard state
we have collected individual dwell measurements with $\Gamma<1.5$. 
For soft state we
have chosen observations with $\Gamma>2.5$ from 2002 year  only. This period was uniquely long  when Cyg X-1
stayed most of the time  in the soft state. 

We also employ a similar procedure to find the appropriate
high-frequency (PCA) part of a broadband PDS. Namely, we 
identify a group of PCA observations by photon spectral index, i.e.
observations in a similar spectral  state and choose the one with PDS 
low-frequency part most closely matching the  appropriate high-frequency part  of EXOSAT PDS 
where they overlap.  
Both RXTE/PCA photon spectra and PDS were
corrected for an effect caused by detector dead time after each event detection.
Deadtime in energy spectra was based on ``The RXTE Cookbook''\footnote{http://heasarc.gsfc.nasa.gov/docs/xte/data\_analysis.html} 
recipe. The details of PDS deadtime correction are described in \citet{rgc00}.  

\subsection{Results of Data analysis and Their Interpretation}
\subsubsection{Power spectrum  evolution vs. photon spectrum evolution}
\begin{figure}[ptbptbptb]
\includegraphics[width=6.in,height=6.5in,angle=-90]{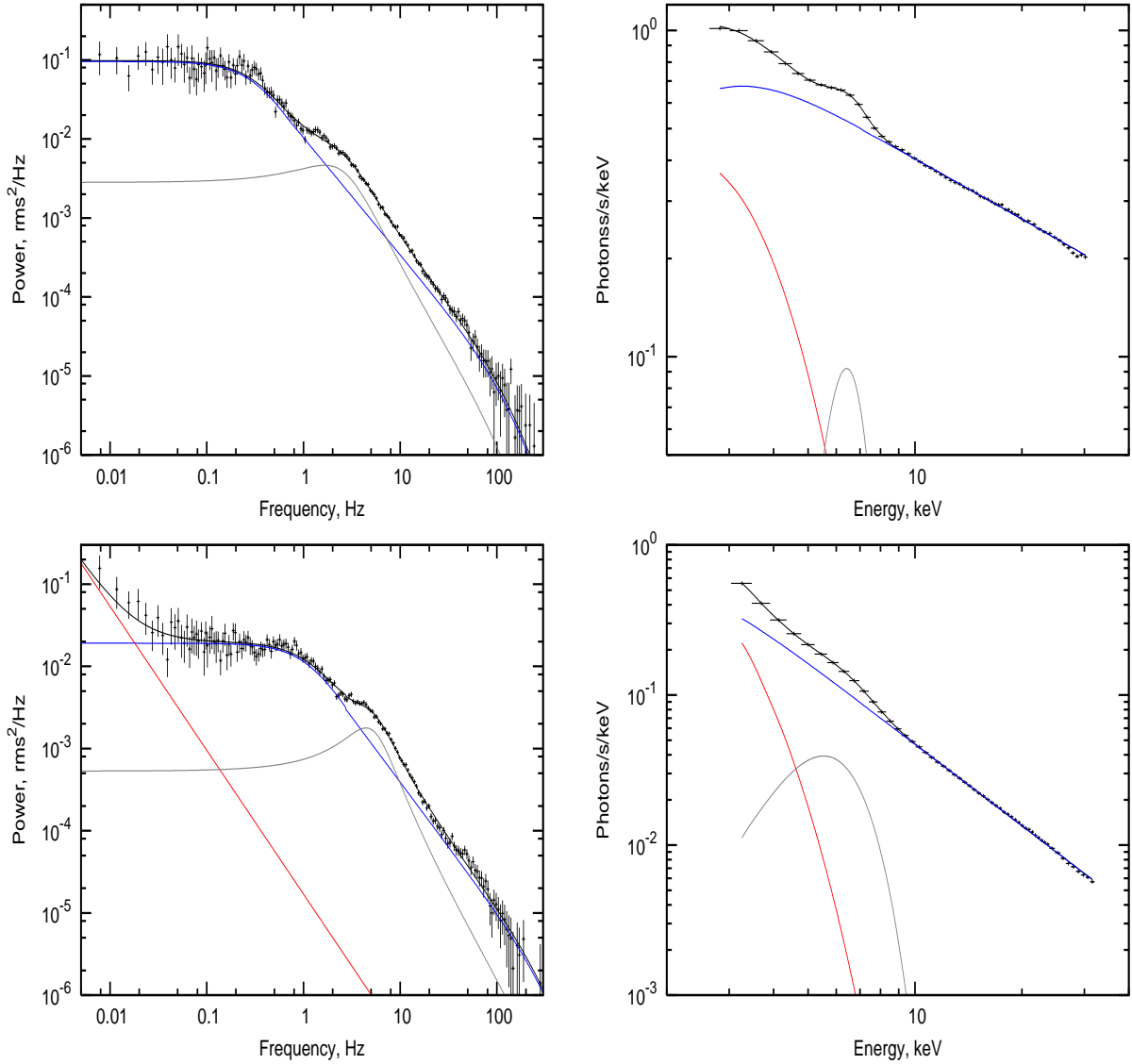}
\caption{Cyg X-1: Observable power spectrum (PDS) (left panel) vs photon spectrum (right panel) 
from observatons 40099-01-05-00(upper panels) and 50119-01-04-01(lower panel). The first
observation is a pure low/hard state with no LF WRN component in the PDS. During the second observation
the source energy spectrum is still hard, but LF WRN is already detectable.
 Data points are shown with error bars. 
 PDS is fitted by a product of the sum of LF and HF WRN power spectra and zero-centered 
Lorentzian (see formula \ref{emerg_PDS_conv}).
We also use  Lorentzians to fit QPO features.  The model fit  $\chi^2$ are shown in Table 1. 
Black line is for  the resulting PDS as red and blue lines present LF and HF components
respectively.  Photon spectrum is fitted by BMC+GAUSSIAN model.
The resulting model spectrum is shown in black, while red and blue curves present thermal and Comptonized 
components respectively. Grey line presents GAUSSIAN of  K$_{\alpha}$ line located at 6.4 keV. 
}
\label{power_lh}
\end{figure}

\begin{figure}[ptbptbptb]
\includegraphics[width=5.in,height=5.5in,angle=-90]{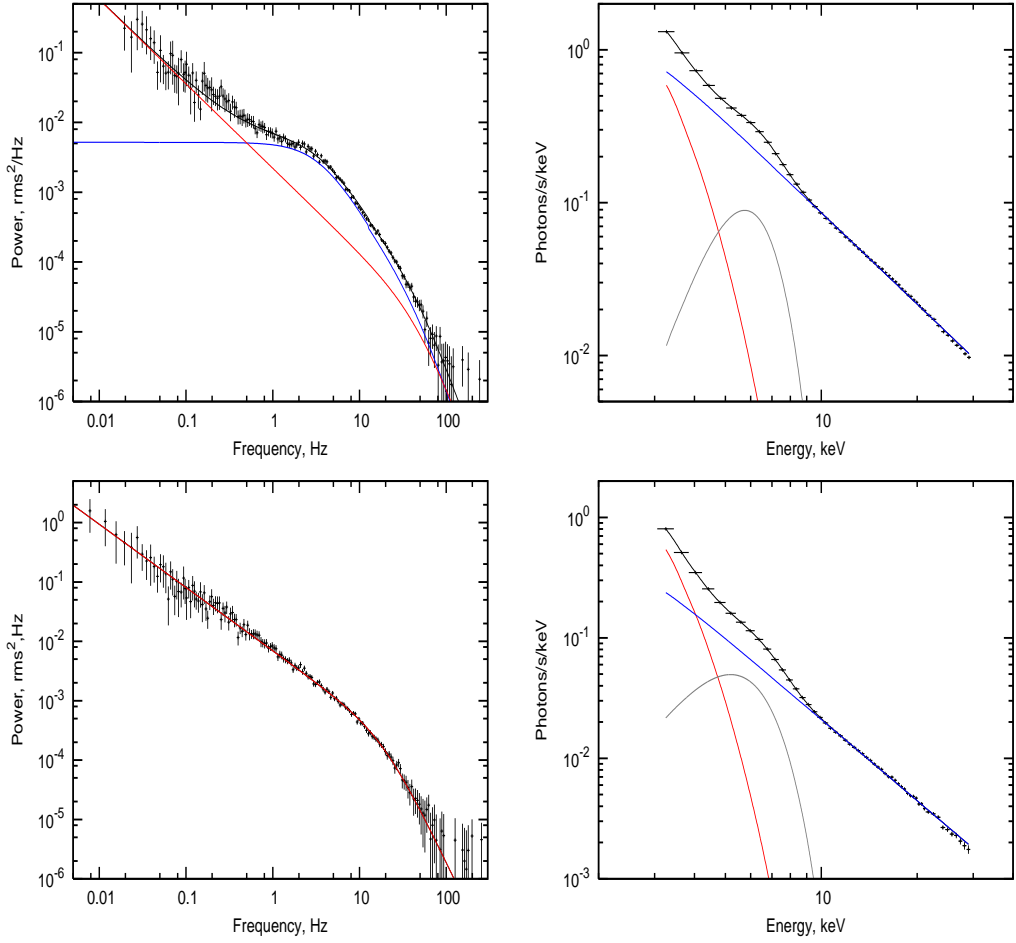}
\caption{The same as Figure \ref{power_lh} but for observations 50109-01-01-00 (upper panels) and
60090-01-14-01(lower panels). The first observation is taken during the intermediate state
just before the transition to high/soft state, which is presented by the second observation.
No HF WRN is present in PDS during high/soft state.}
\label{power_hs}
\end{figure}

Our theoretical model reproduces the observable PDS shape of Cyg X-1 down to 
low frequencies (see Fig. \ref{power_lh} - \ref{power_hs}).  
Note all observational PDSs are presented  in units of rms$^2$/Hz throughout the paper.  
In the low-hard and high-soft states the power spectrum  continuum is fitted by our diffusion  model. 
However, one or two relatively broad Lorentzians  are needed for fitting of QPO features observed in the low-hard  and 
intermediate states of Cyg X-1. 

We clearly see two independent hydrodynamical components  in the accretion flow. 
Their presence are confirmed by power and photon spectra.  They are presumably related to  an extended Keplerian disk 
(Shakura \& Sunyaev 1973) and a compact geometricaly thick sub-Keplerian halo-Compton cloud (see Chakrabarti \& Titarchuk
1995, Narayan \& Yi 1994).
In Figures \ref{power_lh} and \ref{power_hs}  we present the observable evolution of RXTE/PCA PDS and photon spectra of Cyg X-1.
 PDS is fitted by a product of a sum of LF and HF WRN power spectra and a zero-centered Lorentzian 
(see Eqs. \ref{pwsp_varphi_res}, \ref{pwsp_dampos},  \ref{emerg_PDS_conv}-\ref{emerg_PDS_sum})
 plus  the narrow Lorentzians to fit QPO features.   This model is consistent with the data. 
In fact,  we do not see any difference in the fits of  the observed PDSs   if we  use  either a zero-centered Lorentzian or  the driving oscillation Lorentzian of  frequency   $\nu_{dr}$ with quality factor ${\cal Q}\gax1$.  

A black line is for  the resulting PDS as red and blue lines present  the LF component 
$||F_{Y_{in}}(\omega)||^2||F_{G}(\omega)||^2$  and 
HF components $||F_{Y_{out}}(\omega)||^2||F_{G}(\omega)||^2$ respectively.  It is evident that from our fitting procedure  
(see Eq. \ref{emerg_PDS_conv}) that the photon diffusion time as the best-fit parameter $t_{\ast}$ of $||F_{G}(\omega)||^2$ 
 is the same for the LF and HF components of the spectra (see red and blue lines in left panels of Figs.  \ref{power_lh} and \ref{power_hs}).
 We cannot separate them out for each of these individual components. 
 It is worth noting that the model is valid if the variability time scales of the driving oscillations 
 $\varphi_{in}(t)$ and $\varphi_{out}(t)$ are much shorter than the relating diffusion time scales
 of the CC $t_{0,in}$ and extended disk   $t_{0,out }$ respectively.
 The values of the model best-fit parameters for an observational sample are given in Table 1.   

\begin{figure}[ptbptbptb]
\includegraphics[width=3.5 in,height=5.6in, angle=-90]{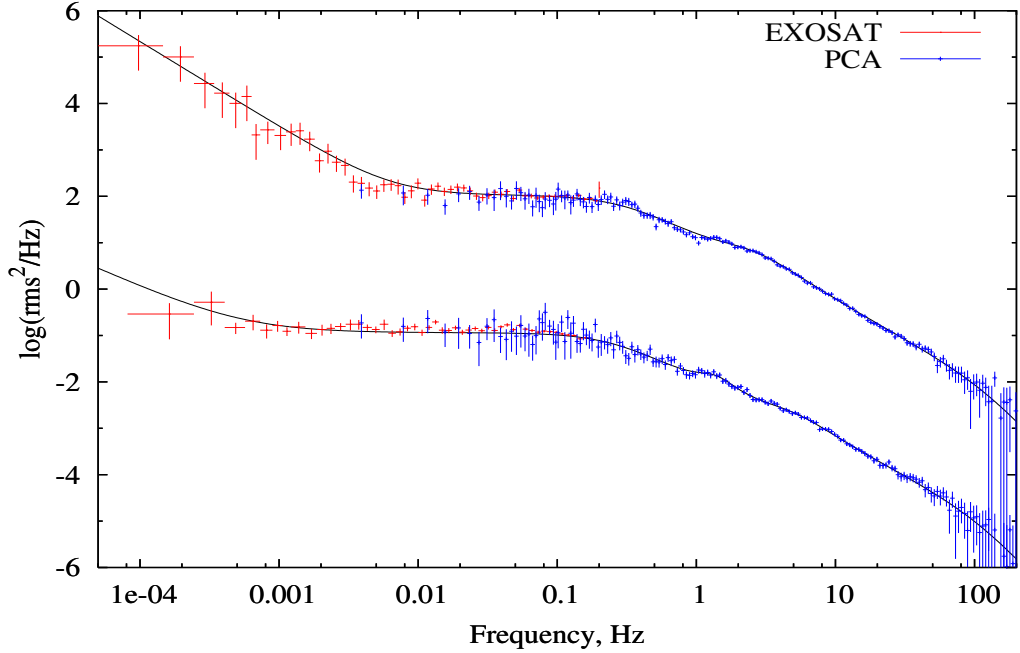}
\caption{Two composite PDSs: EXOSAT spectra with matching high frequency
PCA PDS. PCA data for lower PDS is taken from observation ID 40099-01-03-01 and
for upper PDS - from observation ID 40099-01-05-00. Data are fitted by LF-HF diffusion model:
$\chi^2/N_{dof}=250.1/267=0.94$, $\psi_{CC}=2.32\pm 0.12$, $t_{0,C}=1.8\pm0.3$, $\psi_{D}=2.5$ (fixed) and  $\chi^2/N_{dof}=278.5/267=1.04$, $\psi_{CC}=2.07\pm 0.7$, $t_{0,C}=1.24\pm0.12$, $\psi_{D}=0.3\pm0.3$ (fixed)   for lower and upper panels fits respectively.
 } 
\label{exo_pca_pds}
\end{figure}

We use ASM and EXOSAT data in order to extend the PCA PDSs presented in Figs. \ref{power_lh}-\ref{power_hs}
 to much lower frequencies. In Figure \ref{exo_pca_pds} we show two composite EXOSAT/PCA PDSs of Cyg X-1 for the 
low/hard state.   For  presentation purposes the upper PDSs is multiplied by additional factors of $10^{3}$.
 Note,  the data sets of EXOSAT/ME and RXTE/PCA PDSs were collected at different times and thus to make 
 the composite  EXOSAT/ME and RXTE/PCA PDSs we constructed them by matching low-frequency
part of PCA PDS with high-frequency part of EXOSAT PDS.
 
\begin{figure}[ptbptbptb]
\includegraphics[width=3.5 in,height=5.6in, angle=-90]{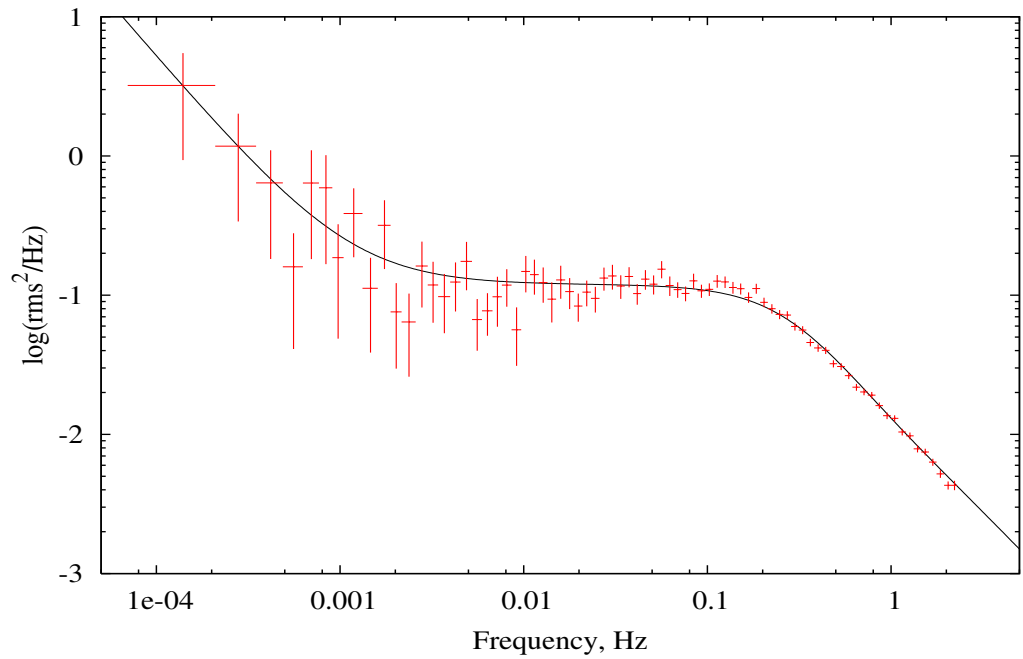}
\caption{High time-resolution EXOSAT/ME PDS of Cyg X-1 from the data collected on 24 July 1984 with high time resolution of ~0.2 sec. Data are fitted by LF-HF diffusion model: $\chi^2/N_{dof}=94.5/74=1.27$,
the best-fit parameters $\psi_{CC}=2.48\pm 0.02$, $\psi_{D}=1.8\pm0.3$, $ t_{0,C}=2.0\pm0.2$ s.
} 
\label{exosat_pds}
\end{figure}

We have plotted EXOSAT power spectrum of 24 July 1984, observation with the best timing
resolution available ~0.2 s,  in Figure \ref{exosat_pds}. We have also  fitted  EXOSAT data
by our LF-HF diffusion model.
No photon diffusion effects were taken into account for this fit as for typical values of
 photon diffusion time $t_*$ are of the order of 1 ms (see Figure \ref{tstar_vs_gamma}) for 
which PDS shape for less than 10 Hz is not affected (see discussion below). 
One can see that
this particular EXOSAT PDS is well described by our LF-HF diffusion
model ($\chi^2=1.27$). It  is worth noting that {\it the low frequency power-law slope is   either barely observed
 or not observed at
 all  in  PCA data alone at low/hard state from 
0.01 Hz to  100 Hz  (Figs.  \ref{power_lh}, \ref{power_hs}) but it
can be clearly seen at longer time scales, probed with EXOSAT}. 

Thus, EXOSAT observes the presence of low frequency power slope related to LF part of our
model along with  the presence of high frequency power slope which
is the HF part of our model.  One can expect that  low
frequency and high frequency power-law slopes can always be observed in the power spectrum of   low/hard
state of Cyg X-1  if one could have simultaneously long observations with
high timing resolution. 

ASM/PCA PDS of the high-soft state is shown on Figure \ref{asm_pca_pds}.
All PDSs shown in Figures  \ref{exo_pca_pds} - \ref{asm_pca_pds} are well fitted by 
our LF-HF component model, a reader can find values of the best-fit parameters and $\chi^2$ for these particular PDSs in the figure captions.  The success of the fitting the data with this additive LF-HF PDS diffusion model (see Eqs. \ref{sum_pds}, \ref{emerg_PDS_sum}) provides   {\it a strong  evidence for the presence of two weakly correlated components in accretion flow of Cyg X-1}.


It is worth noting that the photon diffusion PDS, $||F_{G}(\omega)||^2$  makes a difference in the high frequency part of PDS only.
The photon diffusion PDS is flat for frequencies $\nu\ll 1/(2\pi t_{\ast})$ (see Eq. \ref{TLpower0}). 
In fact, the photon diffusion time scale $t_{\ast}$ 
is  obtained using a high-frequency turnover in PDS (see Figs. \ref{power_lh}-\ref{power_hs})
and  $t_{\ast}$ is related to the photon-crossing time scale 
of the CC-wind (photon diffusion) configuration, $t_{cross}=L/c$, namely 
\begin{equation}
t_{\ast}\sim f\tau_0L/c=f\tau_0 t_{cross} 
\label{t*}
\end{equation}
where $\tau_0=\tau_{CC}+\tau_{W}$ is a sum of
Thomson electron optical depths of Compton cloud $\tau_{CC}$ and wind $\tau_{W}$ (see also Eqs. \ref{cstime}, \ref{tau outflow2}). 

In Figure \ref{tstar_vs_gamma} we show  how  the best-fit parameter $t_{\ast}$ depends on  $\Gamma$.
For $\Gamma> 1.9$ (for intermediate and soft states) the photon diffusion time $t_{\ast}$ correlates with $\Gamma$.
$t_{\ast}$ reaches the highest value about 20 ms in the soft state.
In this state the optical depth of   $\tau_0=\tau_{cor}+\tau_{W}\gax 3$ (see ST06) and thus   
using formula (\ref{t*}) we find  that the photon diffusion size $L\sim ct_{\ast}/(f\tau_0)$ is more than $2\times10^{8}$ cm. 
The distribution of $t_{\ast}$ vs $\Gamma$ is flat for states for which  $\Gamma<1.9$ with some indication of anticorrelation of
$t_{\ast}$ vs $\Gamma$.  The values of $t_{\ast}$ are about  1 ms. 
This anticorrelation is presumably related to the contraction of the Compton cloud when the source only starts progressing 
towards  the soft state and the strong wind is not formed yet (see ST06).
The hard-to-soft spectral transition in Cyg X-1 is likely caused by an increase of the mass accretion rate in the extended disk.
On the other hand, when the mass accretion rate increases the strong wind is launched 
(see Proga 2005 and Appendix E) that leads 
to the rise in the optical depth of the wind 
$\tau_{W}$ and eventually to the correlation of $t_{\ast}$ with $\Gamma$ (see Eq.  \ref{t*}).
We also see the  manifestation of the strong wind development during the spectral
transition as an appearance of the strong iron K$_{\alpha}$ line and disappearance of the QPO features in 
the power spectrum that are washed out in the wind 
(see ST06 and \S 2 for more details).    
  
Photon spectrum is fitted by BMC+GAUSSIAN model.
In Figures   \ref{power_lh}-\ref{power_hs} (right hand panels) the resulting spectrum is shown by black curve  while 
the BMC blackbody and Comptonization  
components are shown by red and blue lines respectively. The gray line presents the Gaussian shape 
of  K$_{\alpha}$ line located at 6.4 keV.
The power of the LF white-red noise component increases with the increase of the power of the BMC BB component 
and strength of K$_{\alpha}$  line (compare Figures \ref{power_lh} and  \ref{power_hs}). 
The HF white-red noise component is shifted to the higher frequencies than that
in the hard state presented in Fig. \ref{power_lh}.  In  the soft state (see Fig. \ref{power_hs},  bottom panel)
the LF white-red noise component (red line) dominates the PDS and the intensity of the BMC BB component reaches the highest value.  

Comparison of the our PDS model  with the data shows that the perturbation (fluctuation) distribution in the accretion flow
is rather smooth. The concentrated fluctuations at the outer edge, $\delta-$function type of fluctuation and their diffusive
 propagation throughout the disk as an origin
of the variable time signal (light curve) is ruled out by the observations. The fluctuations are well distributed over the 
accretion configuration. The  time scale of  the driving fluctuation in the flow  is much shorter than the disk diffusion 
time scale.  This statement is true for all spectral states.  Probably, the steepness of PDS at high frequencies that  occurs 
in all spectral states  can be a sign  of high frequency fluctuations in the accretion flow.  More precise estimate of the 
time scale of the driving oscillation requires further investigation. 

It is important to emphasize that the index of the WRN PDS changes depending on the spectral state.
It is not fixed at the value about one (cf. L97). In fact, the real value of the index (see Eq. \ref{index_pds})
provides us information regarding the viscosity distribution in the accretion flow and ultimately about hydrodynamical 
characteristics of the accretion flow, namely Reynolds number ${\rm Re}$.

\subsubsection{Reynolds number  of the flow and Shakura-Sunyaev disk $\alpha_{\rm SS}-$ parameter as observable quantities}

Using the best-fit parameters of the PDS model we can infer the evolution of the physical parameters   of the source such as 
the disk diffusion time $t_0$, magneto-acoustic QPO frequency $\nu_{MA}$ and  Reynolds number of the accretion flow ${\rm Re}$,  
with the change of  photon index.
We can relate   $t_0$  with ${\rm Re}$ and $\nu_{MA}$ (see Eq. \ref{t0})
\begin{equation}
t_0={4\over{3}}{4\over{(4-\psi)^2}}\left[{{V_{MA}R_0}\over{\hat\nu(R_0)}}\right]\left(\frac{R_0}{V_{MA}}\right)= 
{4\over{3}}{4\over{(4-\psi)^2}}\frac{\rm Re}{a_{MA}\nu_{MA}}.
\label{t0_mod}
\end{equation}
where $a_{MA}$ is a numerical coefficient. 
To relate $V_{MA}/R_0$ ratio with $\nu_{MA}$ we use a formula for magneto-acoustic oscillation frequency derived by Titarchuk, Bradshaw \& Wood (2001), 
hereafter TBW01  (see Eqs. 13, 16 and 17 there):
\begin{equation}
\nu_{MA}= V_{MA}/(a_{MA}R_0)
\label{nu_MA}
\end{equation}
where $a_{AM}\sim2\pi$ is for a pure acoustic case without magnetic field ($\alpha=0$ in Eqs. 13,  17 in TBW01) 
 and $a_{AM}\sim 1$ is for a pure magnetic case ($\alpha=6$  see  Eqs. 13, 16 in TBW01).
The values of $a_{AM}$ presented here are for the free boundary conditions (see TBW01 for details) which are presumably appropriate for the disk-like 
configurations around BHs.  
Formula (\ref{t0_mod}) leads to  equation
\begin{equation}
{\rm Re}=a_{AM}\frac{3}{4}\frac{(4-\psi)^2}{4}(\nu_L t_0)
\label{Re}
\end{equation}
that allows us to  infer a value of $\rm Re$ using the best-fit model parameters $t_0$ and the QPO low frequency $\nu_L$ presumably equals to $\nu_{MA}$.
Ultimately we can find the evolution of ${\rm Re}$ with the photon index $\Gamma$ because the viscosity index $\psi$  and 
the product $\nu_L t_0$ evolves with $\Gamma$ (see Figs. \ref{four_panels}, \ref{t0nuqpo}, Table 1).

In Figure \ref{four_panels} for the HF white-red noise component we present the best-fit diffusion frequency (which is inverse of 
the best-fit diffusion time scale $t_0$)§ vs photon index $\Gamma$ (upper left panel), 
QPO low frequency $\nu_L$ vs $\Gamma$ (right upper panel), the best-fit index of the radial viscosity distribution,  $\psi$  vs 
$\Gamma$ (lower left panel) and the inferred Reynolds number ${\rm Re}$, using $t_0$, $\nu_L$, $\psi$, and  Eq. (\ref{Re}),  vs 
$\Gamma$ (lower right panel).
One can clearly see that  the CC shrinks when the source moves  towards the soft state: QPO frequency $\nu_L$ and inverse 
of the diffusion scale $t_0^{-1}$ increase when the source evolves
to higher indices.  
Also the viscosity ${\hat\nu}\propto R^{\psi/2}$ tends to concentrate to the innermost part of the flow: 
the viscosity index $\psi$ decreases with the photon index. Despite  these correlations of the flow parameters  
the Reynolds number ${\rm Re}$ does not vary much 
within  error bars shown in Fig \ref{four_panels}, namely
${\rm Re} \sim 8\pm 2.5$ (or $\alpha=1/{\rm Re} \sim 0.14\pm 0.04$).

\subsubsection{Diffusion time and QPO frequency}
One can see that our model describes the dynamical behavior of power spectra
with physically meaningful parameters. Indeed, according to the theory (Eq. \ref{t0}), viscous time is proportional 
to $R_0^2$, square of  configuration size (or diffusion frequency $1/t_0\propto1 /R_0^2$)  while QPO
frequency is proportional to $1/R_0$.  As it was already mentioned (see also
ST06) the change in the power index of energy spectra follows the change in
the size of emitting region (CC-wind configuration).  One can see from Figure \ref{four_panels} 
 that the QPO frequency  changes by one order of magnitude   as  the diffusion frequency $1/t_0$ (as the best-fit  parameter
of observed PDS) changes by  two orders of magnitude. This changes correspond to the change in the
 power-law index of photon spectra from 1.5 to 2.0 and therefore to the
change in size of emitting region.  

Another way to demonstrate that Compton cloud contracts  when the source evolves to the softer states is to present a  product  $\nu_{L} t_0$ as a function of $\Gamma$ (see Fig. \ref{t0nuqpo}). $\nu_{L} t_0$ according to our model is a monotonic function of $R_0$  (see Eqs. \ref{t0_mod}-\ref{nu_MA}).
Thus one can conclude using  the inferred dependence of $\nu_{L} t_0$ vs $\Gamma$ that Compton cloud size  $R_0$ really decreases when $\Gamma$ increases  (i.e. when source moves to the softer states).
This behavior of observed QPO and diffusion (time) frequencies is naturally expected in the framework of the our  diffusion model. It proves that our model gives predictive and physically meaningful estimates of
dynamical behavior of the diffusion medium.

\subsubsection{Power-law viscosity index}
We also found that the power-law viscosity index  $\psi$ of the LF frequency component of PDS  depends on the photon index 
$\Gamma$ (see Fig. \ref{psi_gam_soft}).
It increases with $\Gamma$ and then saturates to $\psi=2.9\pm 0.1$ for  $\Gamma> 2.3$. If we assume that the product  
$\nu_L t_0$ is of the order of one
through all states for the  LF frequency component of PDS 
then we obtain that  ${\rm Re_{LF}}\sim 1.5$. In other words the extended disk related to the  LF frequency component 
of PDS (LF WRN) could have higher viscosity 
than that in the CC flow. In fact,  ${\hat\nu}\propto 1/{\rm Re}$ (see Eq. \ref{t0_mod}) but 
the CC ${\rm Re}_{HF}~(\sim8$) is less than the extended disk ${\rm Re}_{LF}~(\sim1.5$). 

\subsubsection{The composite power spectrum of Cyg X-2}


We also constructed the composite PDS for a neutron star source Cyg X-2 using ASM-PCA of RXTE and EXOSAT data 
on Figure \ref{Cygx-2_pds}. 
Cyg X-2 is most of the time in high-soft state, when the photon spectral index is about 4 and higher. 
We found that the broadband  PDS in Cyg X-2 has the structure similar to Cyg X-1. Namely, PDS consists of two  
(LF and HF) components. 
We fit Cyg X-2 PDS using our two components model.  For the LF PDS component the best-fit parameters are:
$t_{D,0}=(6.7\pm1)\times 10^5$ s, $\psi_{D}=1.66\pm0.06$, and for the HF PDS component they are 
$t_{C,0}=0.8056\pm0.0001$ s, $\psi_{CC}=3.11\pm0.02$. QPO lowest frequency is $\nu_L=60.03\pm2.25$ Hz.

TBW01 demonstrated that the magneto-acoustic QPO frequency for a number of NS sources is mostly determined  
by the magnetic (Alfven) frequency $\nu_{M}$.
Thus in  order to calculate the  Reynolds number of the CC flow ${\rm Re}_{CC}$  in Cyg X-2 (NS )
we should  apply formula (\ref{Re}) in which  the numerical factor $a_{AM}\sim 1$ (a pure magnetic case). 
Using the values of the best-fit parameters of the HF component (see above) related to the CC configuration
we obtain that ${\rm Re}_{CC}\sim 7.5$ (or the corresponding $\alpha_{\rm SS}\sim 0.14$). 
It is interesting that for Cyg X-2 the inferred values of ${\rm Re}_{CC}$  and $\alpha_{{\rm SS}, CC}$ 
 are very close to that found in Cyg X-1 (BH). 



\section {Discussion}
The main goal of the presented work is to demonstrate that the emergent timing variability of X-ray emission from compact 
sources is a result of the diffusive propagation of the driving  perturbations which are distributed over the disk. This effect is an 
intrinsic property of any bounded disk-like configuration.
We solved  a problem  of the diffusive propagation of the driving  perturbations (fluctuations). 
We demonstrate that the solution of this problem is reduced to the solution of the initial value problem with 
distributed sources at the initial moment. The formulation is general and classical.
The local driving  fluctuations,  possibly Rayleigh-Taylor local instabilities, gravity waves or  the dynamo as a small-scale stochastic phenomenon,  operating on roughly the local dynamical time-scale,  are  
high-frequency damped quasi-periodic oscillations which frequencies are  related to the local Keplerian frequencies. 
The driving oscillation amplitude is assumed to be a smooth function of the radius.

The basis of the presented power spectrum formation scenario is that  the timing signal is   a result of 
diffusive propagation of driving  perturbations in the bounded configuration (disk or Compton cloud)  
in the same way as X-ray photon spectrum is a result of the photon diffusion  (namely, upscattering of seed photons) in the 
same bounded configuration.
The problem of the diffusive propagation of the  space distribution of high-frequency perturbations is formulated 
as a  problem in terms of  the diffusion equation for  the surface density perturbations. 
This equation is combined with the appropriate boundary conditions (see sections 2, 3).


Our  solution is a convolution of the solution of the initial value problem and the source distribution function (see  \S 2.1 and \S 3 ). 
The solution of the initial value (Cauchy) problem is a linear superposition of exponential shots which are {\it not independent}. 
For example, if the driving perturbations  are distributed  according to the first eigen-function of the diffusion operator 
(see \S 3.4) then the bounded medium works as a filter producing just one exponential shot as a result of the diffusive 
propagation of eigen-function distribution of the seed  perturbations. In the general case the resulting signal is a linear 
superposition of exponential shots which are  {\it related} to the appropriate eigen-functions. 
Furthermore,  in section 2  we demonstrate that  {\it the observed rms-flux relations} (e.g. Uttley, McHardy \& Vaughan 2005)
 is naturally explained by the diffusion solution (model).  In the framework of the linear diffusion theory  the emergent 
perturbations are always linearly related to the driving source perturbations through a convolution of the Green function and 
source distribution (see Appendix B and \S 2.1).

 The asymptotic form of the power spectrum  [Eqs. (\ref{g_pw_sp_lowfr}), (\ref{g_pw_sp_highfr})] is characterized by a flat shoulder (white noise) when the frequency   is less than the inverse diffusion timescale in the disk.
 In other words the  LF WN shoulder    is insensitive to the viscosity law in the disk as a function of radius. 
 The second HF asymptotic form   is  a power law with an  index which is sensitive to the viscosity and perturbation source
 distribution in the disk. When the viscosity {\it linearly} increases with radius and 
 the perturbation sources $f(r)$ are quasi-uniform
  then the index $\alpha$  is exactly $3/2$ ~(see formulas 
 \ref{power_spectrum_y1} and  \ref{pw_sp_highfr}). When the disk viscosity is proportional to 
 $R^{\psi/2}$, the PDS power-law index
 is $\alpha =(3-\delta)/2$, where $\delta=(\psi-2)/(4-\psi)$ ($\delta=0$ for $\psi=2$). 
 We  thus establish that the extended power law ``red noise''  $1/\nu^{(3-\delta)/2}$ 
 is a signature of diffusive propagation of smoothly distributed perturbations in the extended bounded medium (in our case, 
a disk-like configuration, e.g. Keplerian disk or ADAF type   configuration). 
 In other words WRN is a generic consequence of the diffusion theory rather than a specific property of some particular model for the 
variability in accretion disks. It is worth noting this WRN shape  of the PDS continuum is seen for frequencies  
 which are less than the characteristic  frequency of the driving oscillations.   
 We found that the WRN model is consistent with the observations up to  high frequency cutoff  in the PDS,
 which may be an indication of high frequency variability of the driving perturbations in the disk.
The steepness of PDS at high frequencies that  occurs in all spectral states  may  also be   a sign  of high frequency fluctuations 
in the accretion flow and photon diffusion in the CC-wind configuration.

In Figure \ref{driving_qpo} we present two particular examples of PDS (left panels) of two BHC  
GRO J1655-40 and  XTE 1859+226 where the driving QPOs $\nu_{dr}$ are presumably detected by RXTE/PCA.
One can clearly see signatures of $\nu_{dr}$    at $\sim 10- 20$ Hz for  GRO J1655-40 (top) and $\sim185 $ Hz for XTE 1859+226 (bottom) before a high-frequency cut-off.  The continuum of GRO J1655-40's PDS  is well fitted by the white-red noise (WRN) model.
The WRN shape is seen at  all frequencies   which are less than  of $\nu_{dr}$. 
The PDS continuum of XTE 1859+226  is fitted by a sum of LF WRN and HF WRN.  There is a strong QPO low frequency $\nu_L$ at 
$\sim7.5$ Hz near a break frequency of HF WRN.

One can put  a fair question how the rms$^2$ power  of the driving (source) quasi-periodic oscillations is related to 
the observed rms$^2$ (resulting) power which is  a result of the diffusive propagation of the driving perturbations in the disk.
 In \S 2.1 (see Eq. \ref{PL_PX})  we argue that  the resulting power of the emergent signal is  much less than the power of the source (driving) 
perturbations. 
  
The simplest  way to estimate a contribution of the perturbation at a given frequency  to  the total rms$^2$ power
is to construct  a power$\times\nu$ diagram.  In Figure \ref{driving_qpo} we present PDS $\times\nu$ diagrams  of
GRO J1655-40 and  XTE 1859+226 (right top and bottom panels respectively). 
The  rms$^2$ power at  $\nu_{dr}$ of 
order of $10^{-3}$ rms$^2$  is comparable to (GRO J1655-40) or higher (XTE 1859+226) than that at lower frequencies 
(see right panels). The total  rms$^2$ power can be estimated as a sum of power$\times\nu$ over all decades of frequencies. 
 Because  the diffusion (break) frequencies of LF components of GRO J1655-40 and  XTE 1859+226 
presumably occur at frequencies much less than $10^{-2}$ Hz  (they  are not visible in the presented PDSs) 
 we can only guess that the total  power is presumably  of order of  $10^{-2}$  rms$^2$.

\begin{figure}[ptbptbptb]
\includegraphics[width=3.5 in,height=5.6in, angle=-90]{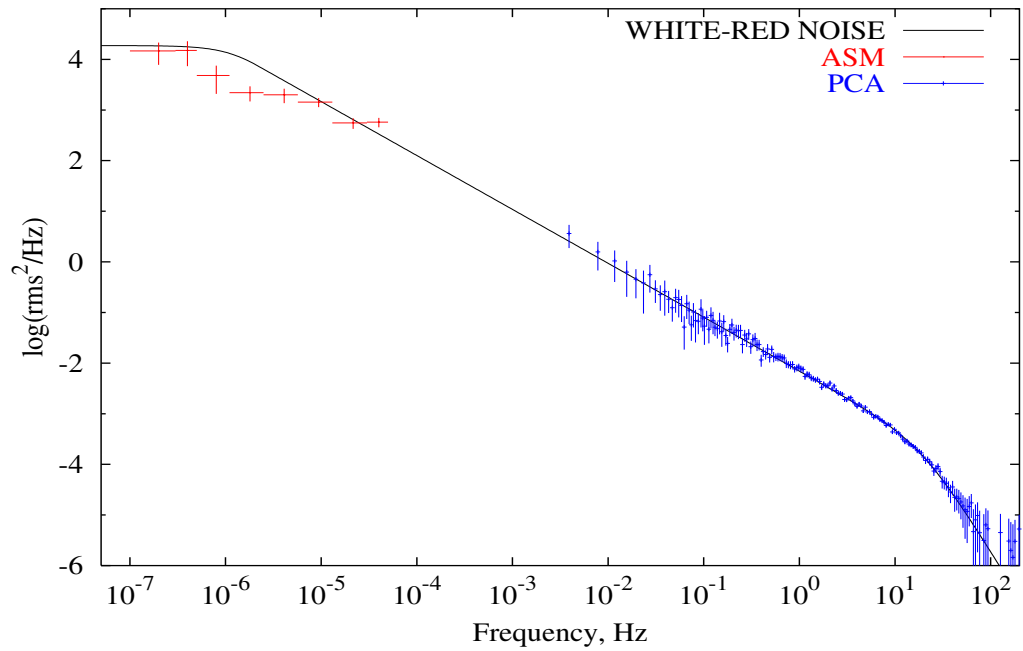}
\caption{  The composite soft state PDS is made by PCA (blue) and ASM (red) PDSs. The PCA PDS is for ObsID 50110-01-52-00 
(see Table 1 and the bottom panel of of Fig. \ref{power_hs}).  Data are fitted by LF-HF diffusion model:
 $\chi^2/N_{dof}=184/228=0.81$, the best -fit parameters $t_{0,D} = (6\pm1.7)\times10^5$ s, $\psi_{D}=2.93\pm0.01$.
}
\label{asm_pca_pds}
\end{figure}

\begin{figure}[ptbptbptb]
\includegraphics[width=5in,height=7.1in,angle=-90]{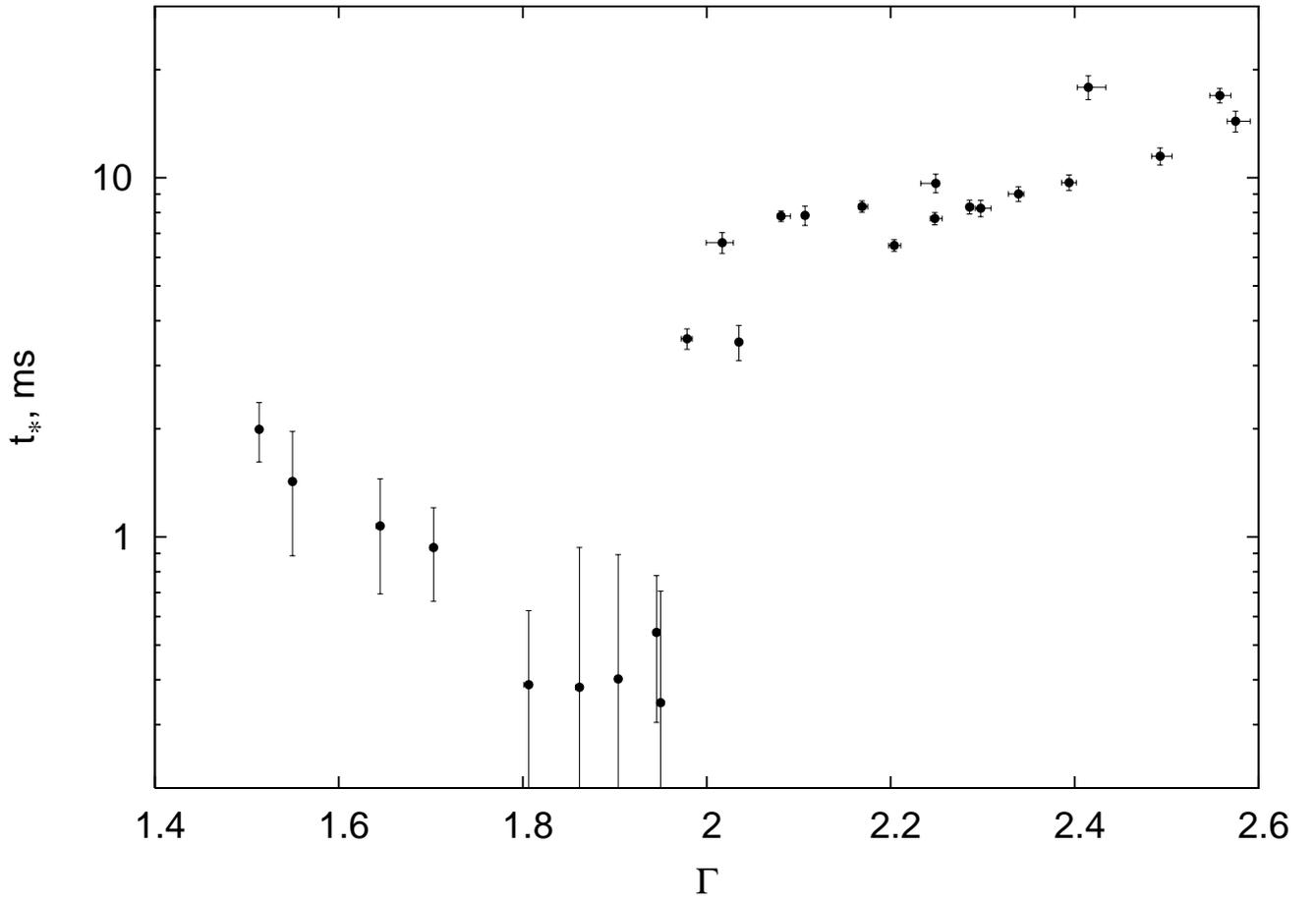}
\caption{ Cyg X-1: HF  photon diffusion PDS (Lorentzian): the best-fit photon diffusion time $t_{\ast}$ vs photon index 
$\Gamma$. 
 }
\label{tstar_vs_gamma}
\end{figure}

\begin{figure}[ptbptbptb]
\includegraphics[width=5in,height=7.1in,angle=-90]{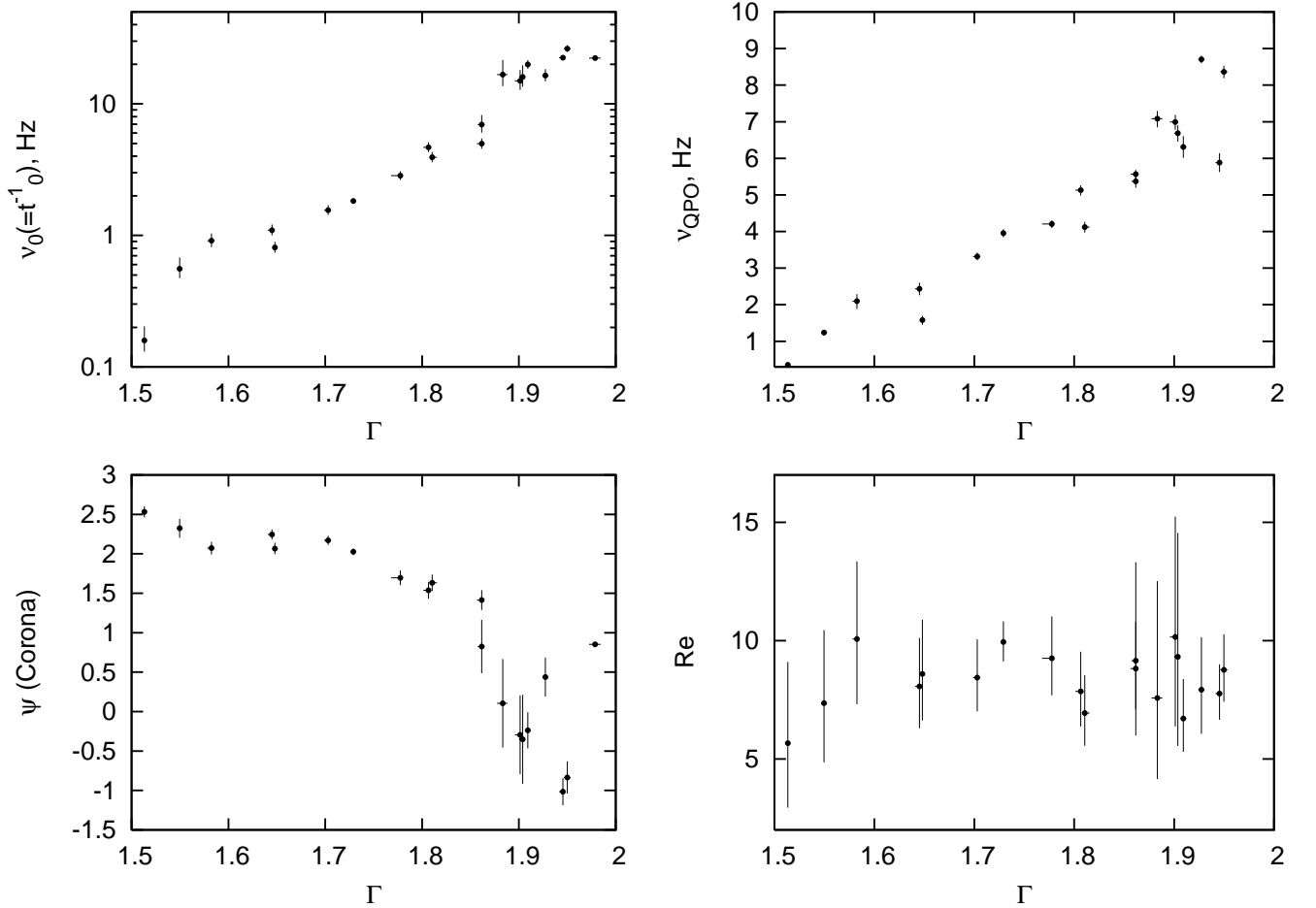}
\caption{Cyg X-1: HF white-red noise component: the best-fit diffusion frequency  $\nu_0=t_0^{-1}$ vs $\Gamma$ (upper left panel), 
QPO low frequency $\nu_{QPO}$  ($\nu_L$) vs $\Gamma$ (right upper panel), the best-fit index of the viscosity distribution $\psi$  vs 
$\Gamma$ (lower left panel) and inferred Reynolds number ${\rm Re}$ (using $t_0$, $\nu_L$, $\psi$, and  Eq.\ref{Re})
 vs $\Gamma$ (lower right panel).}
\label{four_panels}
\end{figure}

\begin{figure}[ptbptbptb]
\includegraphics[width=5in,height=7.1in,angle=-90]{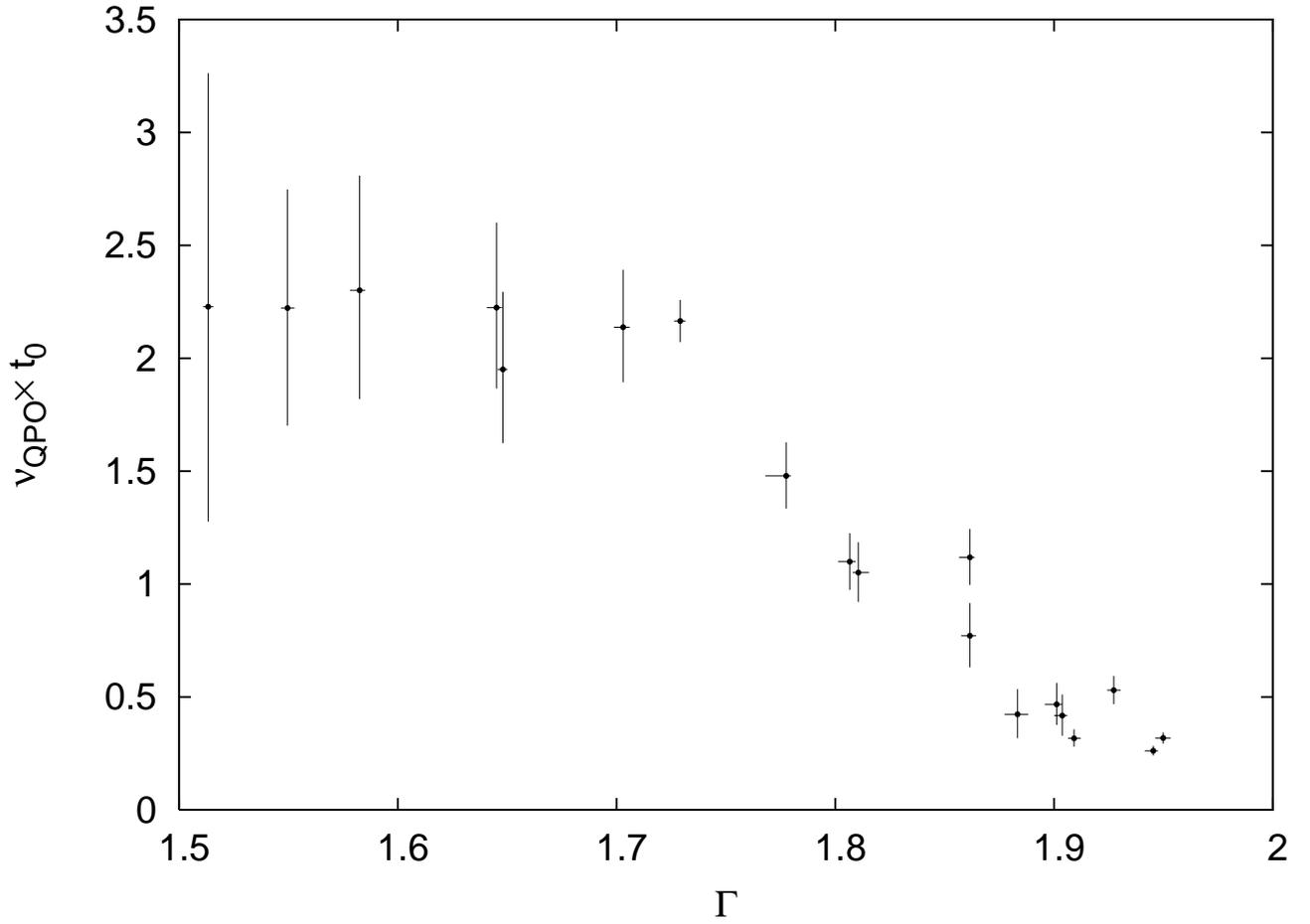}
\caption{Cyg X-1: a product of 
QPO low frequency $\nu_{QPO}$ ($\nu_L$)  and the best-fit diffusion time of HF  WRN $t_0$  vs $\Gamma$.   Decrease of $\nu_{QPO}\times t_0$ with $\Gamma$ implies that Compton cloud contracts  when the source evolves to the softer states.
}
\label{t0nuqpo}
\end{figure}

\begin{figure}[ptbptbptb]
\includegraphics[width=5in,height=7.1in,angle=-90]{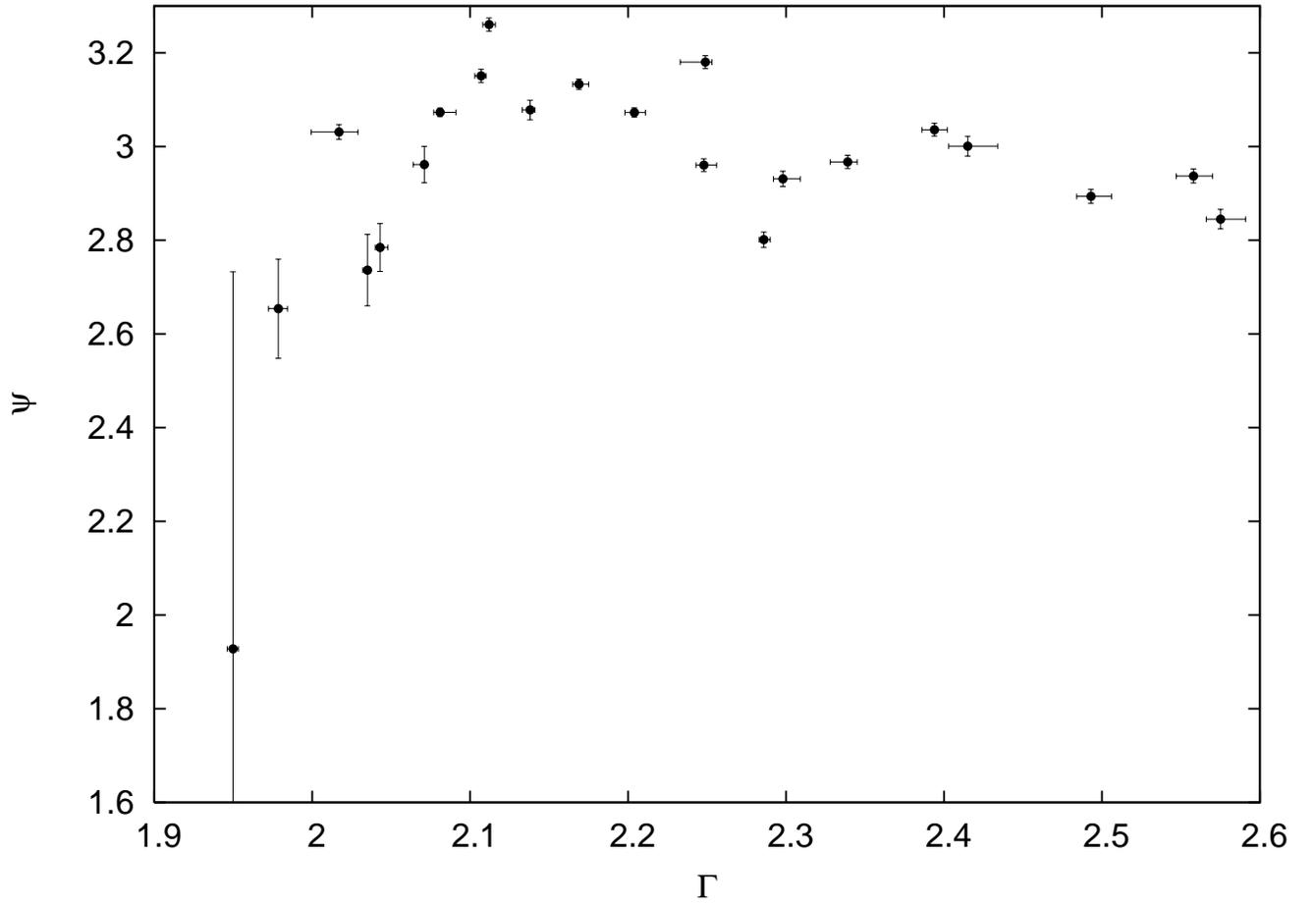}
\caption{Cyg X-1: LF white-red noise component: The best-fit viscosity index $\psi$ vs photon index $\Gamma$.}
\label{psi_gam_soft}
\end{figure}

\begin{figure}[ptbptbptb]
\includegraphics[width=5in,height=7.1in,angle=-90]{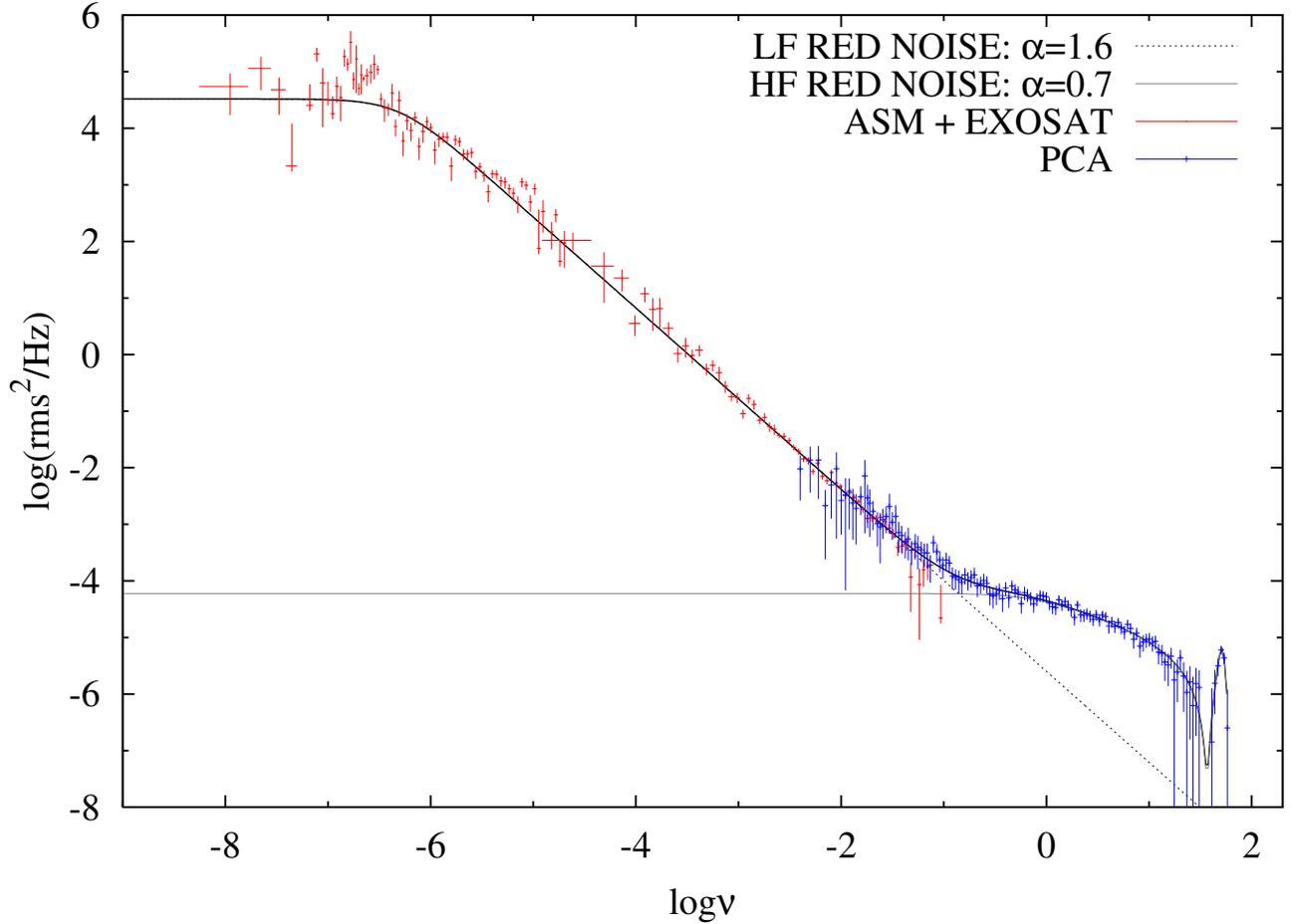}
\caption{ EXOSAT-ASM-PCA (RXTE) power spectrum of  Cyg X-2 in frequency range that covers 10 orders of magnitude.
One can clearly see  low and high frequency (LF and HF) white-red noise components in PDS, related to the extended Keplerian
disk and relatively compact, innner disk-like configuration (Sub Keplerian Compton cloud) respectively. 
Each of these two components is perfectly fitted by our white-red noise model (see Eqs. \ref{emerg_PDS_conv} and \ref{emerg_PDS_sum}), dotted and  solid lines are for LF and HF best-fit models respectively, $\chi^2/N_{dof}=393.2/244=1.6$
(see the text for the best-fit parameters values).
  }
\label{Cygx-2_pds}
\end{figure}

\begin{figure}[ptbptbptb]
\includegraphics[width=5in,height=7.1in,angle=-90]{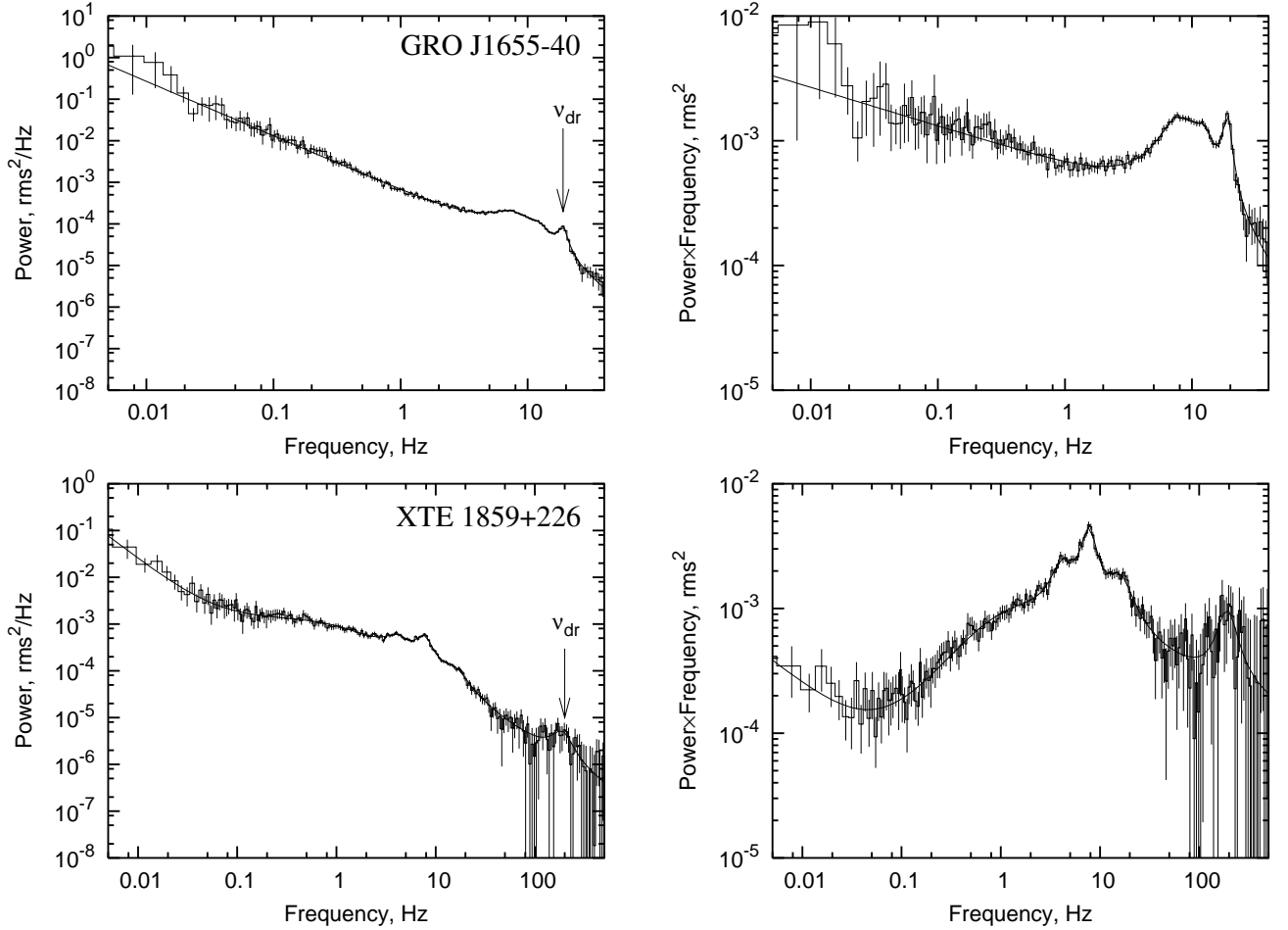}
\caption{ RXTE/PCA power spectra (left panels) and power$\times$frequency diagrams (right panels)  of  GRO J1655-40 (top) and XTE 1859+226 (bottom). 
One can clearly see  QPO frequencies $\nu_{dr}$  
at $\sim 10-20$ Hz for  GRO J1655-40 and $\sim185 $ Hz for XTE 1859+226 before a high-frequency cut-off. The  rms$^2$ power at  $\nu_{dr}$ is comparable (GRO J1655-40) or higher (XTE 1859+226) than that at low frequencies (see right panels).    
 }
\label{driving_qpo}
\end{figure}

On the other hand  in the composite ASM/PCA power spectrum of Cyg X-1 (see Fig. \ref{asm_pca_pds})  the break frequency of LF component is clearly identified  at about $10^{-6}$ Hz. 
 It should be  noted that the RN high-frequency tail  has a cutoff    around 10 Hz which can be an indication of the driving perturbation frequency, i.e. $\nu_{dr}\sim 10$ Hz.
 Thus we can calculate the total  
rms$^2$ power by a direct integration of the power spectrum.
In order   to calculate this power  
one should integrate the red noise (RN) component of the resulting PDS  
$||F_x(\nu)||^2_{\nu,RN} \sim10^{4}(\nu/10^{-6}~{\rm Hz})^{-1.06} {\rm rms}^2/{\rm Hz}$ and the white noise (WN) component 
$||F_x(\nu)||^2_{\nu,WN}\sim10^{4}~{\rm rms}^2/{\rm Hz}$. It  
is obvious that the RN component power $P_{RN}=
0.44~ {\rm rms}^2$. The integration of the WN 
 PDS component  from 0 to $10^{-6}$ Hz gives us that the WN component power  $P_{WN}=
 10^{-2}{\rm rms}^2$.
 Thus the total power $P_{tot}=P_{RN}+ P_{WN}\approx0.45~{\rm rms}^2$.  
It is easy to check that the similar estimate can be obtained using multiplication of this PDS 
by a frequency $\nu$ followed by summation   of a power$\times\nu$ over frequency decades. 

We remind a reader that the resulting power spectrum is a product of WRN and driving oscillation power spectra (see formulas \ref{pwsp_varphi_res} and
\ref{pwsp_product}).   An interesting question is what is a relative contribution  of the WRN PDS and the driving oscillation PDS in the resulting PDS of Cyg X-1.

In fact,  the  WRN PDS  is  normalized to $1/(Dt_0)$ where $D\gax1$ (see Eqs. \ref{FY_generalA}).  
The diffusion time scale in the disk  $t_0$ as the best-fit parameter of WRN PDS in Cyg X-1 equals to 
$6\times10^{5}$ s (see Fig. \ref{asm_pca_pds}). 
Thus in this case of Cyg X-1 the  WRN PDS  should be presented  as followed 
\begin{equation}
||F_{Y}(\nu))||^2_{\nu}\approx A_N \left\{\begin{array}{ll}
1 & \mbox{if $\nu\leq 10^{-6}$ {\rm Hz}}\\
(\nu/10^{-6}{\rm Hz})^{-1.06} & \mbox{if $\nu\geq 10^{-6}$ Hz}
\end{array}\right.
\label{WRN_norm}
 \end{equation} 
 where $A_N\approx0.1$ is a normalization constant.
 
 Using the  total power of the driving oscillations $P_{dr}$ 
 we can rewrite  the driving oscillation PDS as (see Eqs. \ref{pwsp_dampos} and \ref{Four_phi}) 
 \begin{equation}
||F_{\varphi}(\nu)||_{\nu}^2=\frac{\hat\Gamma_{dr}  P_{dr}/(a{\pi})}{(\nu-\nu_{dr})^2 + (\hat\Gamma_{dr}/2)^{2}}
 \label{driv_norm}
 \end{equation}
where $\hat \Gamma_{dr}$ is a full width of half maximum (FWHM) of the Lorentzian and a constant $a$ varies in the range between 
$1$ and $2$ depending on the ratio of $2\nu_{dr}/\hat\Gamma_{dr}$ (for example $a=1$ and $a=2$ when 
$2\nu_{dr}/\hat\Gamma_{dr}\ll 1$ and $2\nu_{dr}/\hat\Gamma_{dr}\gg1$ respectively). 
 At frequencies $\nu\ll\nu_{dr}$ the driving PDS is a constant
 \begin{equation}
||F_{\varphi}(0)||_{\nu}^2=|| F_{\varphi}(0)||_{\nu}^2=\frac{\hat\Gamma_{dr}  P_{dr}/(a\pi)}{\nu_{dr}^2 + (\hat\Gamma_{dr}/2)^{2}}.
 \label{driv_norm0}
 \end{equation}
Because for any power spectrum $||F(\omega)||^2$ 
$$|| F(\omega)||^2d\omega=|| F(2\pi\nu)||^22\pi d\nu=||F(\nu)||_{\nu}^2d{\nu}$$ 
we have that (compare with Eq. \ref{pwsp_varphi_res})
\begin{equation}
|| F_x(\nu)||_{\nu}^2=(2\pi)^{-1} ||F_{\varphi}(\nu)||_{\nu}^2||F_{Y}(\nu)||_{\nu}^2.
 \label{Fnu_product}
 \end{equation}
 From  Eqs. (\ref{WRN_norm}), (\ref{driv_norm0}), (\ref{Fnu_product})
we obtain that  for the ASM/PCA power spectrum of Cyg X-1 
 \begin{equation}
||F_{\varphi}(0)||_{\nu}^2=2\pi||F_{x}(0)||^2/||F_{Y}(0))||^2=2\pi\times10^{4}/0.1~ ({\rm rms}^2/{\rm Hz})\sim6\times 10^5~
({\rm rms}^2/{\rm Hz}).
 \label{driv_cygx-1}
 \end{equation}
As an application of the preceding analysis 
we   can conclude that $P_{dr}\gax6.3\times10^6$ rms$^2$   because $P_{dr}\gax\nu_{dr} 
||F_{\varphi}(0)||_{\nu}^2$. If we compare this with the integrated power  $P_x=P_{tot}=0.45$ 
 for the Cyg X-1 PDS (see Fig. \ref{asm_pca_pds}) we find that the integrated power of the driving oscillations $P_{dr}$ is higher  by 7 orders of magnitude than that for the emergent signal, namely
 ($P_{dr}/P_x)_{CygX-1}\gax 1.4\times10^7$.  In other words {\it the resulting observed integrated power is very small fraction ($\sim10^{-7}$)  of the integrated power of the entire disk}\footnote{The referee  was first who inferred   this ratio using the Cyg X-1 PDS presented in Fig. \ref{asm_pca_pds}.} .   This is precisely what was predicted by our analysis of the  diffusion model (see Eq. \ref{PL_PX}).
In fact, the theoretically predicted ratio
\begin{equation}
\left(\frac{P_{x}}{P_{dr}}\right)_{diff}\sim\frac{1}{2\pi\nu_{dr}t_0{\cal Q}D}=\frac{1}{(2\pi \times10{\rm Hz})~(6\times10^5{\rm s}){\cal Q}D}=2.6\times10^{-7}/({\cal Q}D)
\label{theory_obs}
\end{equation}  
is comparable with that inferred  for the observed Cyg X-1 PDS. 
Namely
\begin{equation}
\frac{(P_{x}/P_{dr})_{CygX-1}}{(P_{x}/P_{dr})_{diff}}\sim0.5 {\cal Q}D\sim1
\label{theory_obsm}
\end{equation}
where a factor ${\cal Q}D\gax1$.
 
 We emphasize that the presented  theory (model)  provides  new  physical insights into accretion processes 
that occur around compact objects:
\begin {enumerate}
\item We shed light on the nature of the power spectrum continuum. The PDS shape is a white-red noise which is a result of diffusive 
propagation of high-frequency damped quasi-periodic driving oscillations.
The shape is determined by the disk diffusion time scale and the viscosity distribution in the disk.

\item The ``rms-flux relation''  found in the observation [see e.g. Uttley (2004) and Uttley et al. 2005] is naturally 
explained in the framework of the diffusive propagation of the disk perturbations.

\item The decay of the variability power within 0.1-10 Hz frequency range  during the source evolution from the hard to  soft states is well known observational phenomenon. 
We found that this effect is a natural consequence of the diffusive 
propagation of  high-frequency driving oscillations. 
We demonstrate, using the diffusion theory and observed Cyg X-1 PDS,  that the integrated power of the resulting PDS $P_x$  is only a tiny fraction of the integrated power of the driving oscillations $P_{dr}$  distributed over the disk. 
Furthermore, we demonstrate that the resulting power $P_x$ is inversely proportional to the characteristic frequency of the driving oscillations $\nu_{dr}$ which is presumably scaled with the frequency of the local gravity waves in the disk (Keplerian frequency).  
When the source evolves to softer state, the Compton cloud region 
becomes more compact (see ST06)
and thus the power  $P_x$   declines towards soft states. 
\item To fit the observed PDSs we use WRN model for the continuum and Lorentzians to determine  the various QPO frequencies 
seen as bumps in PDS.  As a result of this fitting, in particular, we found the diffusion time of Compton cloud disk-like 
configuration $t_0$ and the low QPO frequency (see Fig. \ref{four_panels} in the paper) is a function of photon index. It is seen from 
 Fig. \ref{four_panels} that as QPO frequency changes by one order the  inverse of $t_0$ (a diffusion frequency) changes by almost two 
orders of magnitude in a given photon index range. This is a real self-consistency check of the diffusion theory vs 
observations because QPO frequency $\nu_{L}$ is proportional to 
$1/R_0$ and $1/t_0$ is proportional to $(1/R_0)^2$ ($R_0$ is the Compton cloud size). 

\item   The  rising QPO frequency and decreasing $t_{C,0}$  with the spectral index $\Gamma$  is a signature of  
the  Compton cloud   contraction  as it progresses towards the soft state.  QPO frequencies varies inversely  
with the size of the region.   (see ST06 for details of  spectral and timing analysis of
X-ray data for Cyg X-1 collected with the {\it  RXTE}).
 
\item The best-fit  model parameters $ \nu_{L}$ and $t_0$ allow us to determine Reynolds numbers Re directly from the 
observed power spectra. Thus, our model provide an opportunity to study the magnetohydrodynamics of the accretion flow directly  by using the observed PDS. We found that Re has a constant  physical value  (within the error bar range)  as a  function of several varying parameters  (diffusion time, QPO frequency and the WRN power-law index). 
\end{enumerate}


\section{ The main results and conclusions}

We conclude by summarizing the main results of the presented diffusion theory.

We have presented a detailed mathematical analysis of the perturbation diffusive
propagation.We investigated the intrinsic
properties of the disk density evolution equation (\ref{main_eq}) with the appropriate boundary and initial conditions in a general
case.  We have analyzed the diffusion models determined by
the disk  viscosity dependence on the radius for various perturbation sources in the disk.  
We have examined the case where the viscosity is a power law function of position in the disk.  
Using the perturbations of the  disk surface density $\Delta \Sigma (r,t)$ we are able to infer
the evolution of the perturbations of the mass accretion rate in the inner disk edge and ultimately the 
perturbations of  the X-ray luminosity as a function of time, $\Delta L_x (t)$.
Then we calculate the power spectrum using the Fourier transforms of $\Delta L_x (t)$ and the driving perturbations.    
The PDS continuum (White-Red-Noise) is a   a power spectrum of the  diffusion response of the disk-like configuration to 
the high frequency (local) driving disk oscillations. Whereas X-ray photon spectrum is the result of the soft photon diffusion 
upscattering (Comptonization) in the disk-like configuration (Compton cloud),   the PDS is formed in the 
same configuration as a result of the diffusive propagation of high-frequency local driving perturbations.   
 This solution is robust and generic.  The result can find many applications 
where oscillations of  boundary configurations are studied. Here we apply this solution to the particular astrophysical case of the  disk oscillations.

The resulting model time signal as a linear combination of quite a few {\it related} exponential shots  is in a good 
agreement with the observations. 
 The observable PDS is perfectly fitted by a sum of LF and HF white-red noise power spectra.
 {\it This fact can be interpreted as an observational evidence of the presence of two independent components in the accretion flow.}
 One is related to the extended geometrically thin disk (LF PDS component) and the other  - to the geometrically 
thick compact configuration (HF PDS component). Each of the white-red noise (WRN) components has two free parameters,  
the diffusion time scale $t_0$  and the viscosity 
index $\psi$. The value of the parameter $t_0$ has a physically plausible value for the
viscous timescale of the disk-like configuration and ultimately is applied to the
calculation of the Reynolds number of the accretion flow ${\rm Re}$. In fact, the  value of ${\rm Re}$ can be 
 inferred  if $t_0$, $\psi$ and QPO low frequency $\nu_L$ are known from observations (see Eq. \ref{Re}).

 In the observed power spectra (particularly in the intermediate state)
we deal with two diffusion time scales, one $t_{C,0}$ is related to the inner compact region, presumably 
Compton cloud (sub-Keplerian disk, ADAF) 
and the other  $t_{D,0}$  is related to a much larger disk. The diffusion time scale of the inner region is scaled with the mass 
of the central object  and  $t_{D,0}$ is scaled with the orbital period of the system. 
The  diffusion time scale $t_{\ast}$ (as the best-fit parameter of the Lorentzian)
determines a high-frequency turnover in PDS (see Fig. \ref{power_lh}). It  may  be related to the combined effect of the 
photon diffusion in the CC-wind configuration and  high frequency (local) fluctuations in the accretion flow. 
More precise estimate of the time scale of the driving oscillation requires further investigation.


We present the broadband PDSs of black hole source Cyg X-1 in hard (Fig. \ref{exo_pca_pds}) and soft (Fig. \ref{asm_pca_pds}) states,
illustrating the presence of  LF and HF components in hards state and the absence 
 of HF component ((or a weak HF) in the soft state.
In Figures  \ref{power_lh}, \ref{power_hs} we show the evolution of LF and HF components in high frequency PDS of Cyg X-1. 
 We also find that Cyg X-2 PDS also consists of LF and HF components (see Figure \ref{Cygx-2_pds}). 
The best-parameters of the model allows us to determine 
the diffusion time scales of geometrically thin extended disk $t_{D,0}$ and geometrically thick configuration (Compton cloud) $t_{C,0}$.
 They  differ by almost six orders of magnitude, namely $t_{D,0}\sim 0.7\times 10^6$ s  and $t_{C,0}\sim0.8 $ s.
We infer the Reynolds numbers ${\rm Re}$  and related $\alpha_{\rm SS}$ parameters of the CC flow  for Cyg X-1 and Cyg X-2. 
We find they are quite similar for each source, namely ${\rm Re}$ is about 7.5  and  $\alpha_{\rm SS}$  is about 0.14.






LT  appreciates productive questions by Marat Gilfanov, Ralph Fiorito,  Martin Laming, and Demos Kazanas. 
VA acknowledges  useful discussion with Marat Gilfanov and a partial support of this work from the program ``the origin 
and evolution of stars and galaxies" of the Russian Academy of Science. NS thanks Craig Markwardt for outstanding 
software products. We also acknowledge the referee's contribution in Discussion section.

\newpage

\appendix
\section{Power spectrum of weakly correlated signals}
If two signals $\Delta L_1(t)$ and $\Delta L_2(t)$ in the system  are weakly correlated then  the resulting power spectrum  is a sum of the corresponding spectra 
Namely, 
\begin{equation}
\int_0^{\infty}[\Delta L_1(t)+\Delta L_2(t)]^2dt \approx\int_0^{\infty}[\Delta L_1^2(t)+\Delta L_2^2(t)]dt
\label{sum_pdsm}
\end{equation}
Using Parseval's theorem for Fourier transforms, we obtain that 
\begin{equation}
\int_0^{\infty}[\Delta L_1^2(t)+\Delta L_2^2(t)]dt=\int_0^{\infty}[||F_{x_1}(\omega)||^2+||F_{x_2}(\omega)||^2]d\omega.
\label{sum_energy}
\end{equation}
where  $F_{x_1}(\omega)$ and $F_{x_2}(\omega)$ are Fourier transforms of $\Delta L_1(t)$ and 
$\Delta L_2(t)$ respectively.
Thus the energy per unit frequency, resulting power spectrum,  $||F_{x_1+x_2}(\omega)||$ is a sum of two components
\begin{equation}
||F_{x}(\omega)||=||F_{x_1}(\omega)||^2+||F_{x_2}(\omega)||^2.
\label{sum_pds_2}
\end{equation}

\section{Diffusion of the driving perturbations. Mathematical details}
\subsection{Solution of Eq. (\ref{dif_driv_eq}) for the general case of the spatial and time dependent distribution  of driving perturbation}
The diffusion equation for the time variable quantity $W(R,t)$,  related to the surface density  can be written  as    (compare with Eq. \ref{dif_driv_eq}):
\begin{equation}
{{\partial W}\over{\partial t}}={\bf\Lambda_R} W+\Phi(t,R).
\label{general_dif_driv_eq}
\end{equation}

The solution of equation (\ref{general_dif_driv_eq}) with homogeneous initial condition $W(R,0)=0$
can be presented using the Green's function $G(R, \xi, t)$ derived  by W01 (see Eq. 21 there), 
namely, $G(R, \xi, t)=V_{x_0}(x,t)/ [\nu (x_0)/\pi]$ (where $x=R^{1/2}$ and $ x_{0}=\xi^{1/2}$) . Thus the solution is (compare with Eq. \ref{convd})
\begin{equation}
W(R,t)=\int_0^t \int_{R_{in}}^{R_0} G(R,\xi, t-t^{\prime})\Phi(t^{\prime},\xi)d\xi dt^{\prime}.
\label{general_convd}
\end{equation} 
Then a  Fourier transform of $W(R,t)$ is 
\begin{equation}
F_{W}(R, \omega)=\int_{R_{in}}^{R_0} F_{G}(R,\xi, \omega)F_{\Phi}(\xi, \omega)d\xi.
\label{Four_convd}
\end{equation} 
Without any loss of generality  we assume 
that  the local  driving perturbations are  damped quasiperiodic oscillations, namely
\begin{equation}
\Phi(t,\xi)=A_{\Phi}f(\xi)\exp(-\frac{1}{2}\Gamma_{dr} t+i\omega_{dr}t)
\label{driving_fourier}
\end{equation}
where $\omega_{dr}$ and the damping factor are in principle functions of $\xi$, $A_{\Phi}$ is a numerical factor related to an amplitude  of the driving  quasiperiodic oscillations.
In this case a Fourier transform of $\Phi(t,\xi)$ is
\begin{equation}
F_{\Phi}(\xi, \omega)=\frac{A_{\Phi}f(\xi)}{\Gamma_{dr}/2+i(\omega-\omega_{dr})}.
 \label{Four_phi}
\end{equation} 
For a product of continuous functions $g_1(\xi)$,  $g_2(\xi)$ and $g_3(\xi)$ 
 the mean value theorem states that 
  \begin{equation}
\int_{a}^{b}g_1(\xi)g_2(\xi) g_3(\xi) d\xi= g_3(\xi_{\ast})\int_{a}^{b}g_1(\xi)g_2(\xi) d\xi
\label{mean_theorem}
\end{equation} 
where $\xi_{\ast}$  is between $a$ and $b$.
Using this mean value theorem and formula (\ref{Four_phi}) for $F_{\Phi}(\xi, \omega)$ we can present 
Eq. (\ref{Four_convd}) as  
 \begin{equation}
F_{W}(R, \omega)=\frac{A_{\Phi}}{\Gamma_{dr}(\xi_{\ast})/2+i[\omega-\omega_{dr}(\xi_{\ast})]}
\int_{R_{in}}^{R_0} F_{G}(R,\xi, \omega)f(\xi)d\xi
\label{mean_Four_convd}
\end{equation} 
where $\xi_{\ast}$ is some mean value of $\xi$ between $R_{in}$ and $R_0$.
The integral in the right hand side of equation (\ref{mean_Four_convd}) is the Fourier transform of the solution of  the intitial value problem $X(\xi,t)$  for the homogeneous  diffusion  equation (\ref {hom_dif_driv_eq}) 
with the initial function $X(\xi,0)=f(\xi)$:
 \begin{equation}
F_{W}(R, \omega)=\frac{1}{\Gamma_{dr}(\xi_{\ast})/2+i[\omega-\omega_{dr}(\xi_{\ast})]}
F_X(R,\omega).
\label{modmean_Four_convd}
\end{equation}  
Using Eqs. (\ref{modmean_Four_convd}), (\ref{convdL}), (\ref{signal_y}) we obtain the Fourier transform  of the resulting X-ray signal due to the diffusion of the driving perturbations 
\begin{equation}
F_{x}(\omega)=3\pi\varepsilon_{eff}
\frac{\partial [x{\hat \nu} F_{W}(\omega,x^2)]}{\partial x}|_{x=0}
\label{PDS_genA}
\end{equation}

Then the  power spectrum $||F_{x}(\omega)||^2$ is   a product 
  \begin{equation}
 ||F_{x}(\omega)||^2=\frac{A_{\Phi}^2||F_Y(\omega)||^2}{[\omega-\omega_{dr}(\xi_{\ast})]^2 +[\Gamma_{dr}(\xi_{\ast})/2]^{2}}
\label{pwsp_product}
\end{equation}  
where $Y(t)$ is determined by formula  (\ref{signal_y}).
Comparison of formulas (\ref{pwsp_dampos}) and (\ref{pwsp_product}) leads to the conclusion that the value of
 $\omega_{dr}$  in formula  (\ref{pwsp_dampos}) is some mean value of the rotational frequency of the local quasiperiodic oscillations.  This frequency, likely related to the local Keplerian  frequency  is much higher 
 than the diffusion frequency of the disk-like configuration (see \S  3). 

\subsection{A relation between the total power  of the driving (source) oscillations and resulting power due to diffusion of the driving perturbations}
 We can estimate  the integrated power  of $ ||F_{x}(\omega)||^2$ if  we estimate that for 
 $ ||F_{dr}(\omega)||^2$ and $ ||F_{Y}(\omega)||^2$.
But from  Parseval 's theorem of Fourier transform we know that 
\begin{equation}
\int_0^{\infty}  ||F_{Y}(\omega)||^2dt=\int_0^{\infty}Y^2(t)dt.
\label{Y2_normA}
\end{equation}
 The integral of $Y^2(t)$  in Eq. (\ref{Y2_normA}) can be evaluated using the integral of $Y(t)$. 
It is easy to show  that   
\begin{equation}
\int_0^{\infty}Y(t)dt=\varepsilon_{eff} \int_0^{R_0}f(R)2\pi RdR=1.
\label{Y_normA}
\end{equation}
 To derive Eq. (\ref{Y_normA}) we integrate Eq. (\ref{hom_dif_driv_eqm}) over t (from $0$ to $\infty$) and over x (from $0$ to $x_0$) and we also use  
 Eqs. (\ref{init_cond0mm}),  (\ref{C_S}) and  the outer boundary condition $\partial {\cal X}/{\partial x}(x_0,t)=0$ (see  W01).

When the driving perturbations are distributed in the disk according to the first eigenfunction of the space  operator of the diffusion equation (\ref{dif_driv_eq}) then $Y(t)= {t_\ast}\exp(-t/t_{\ast})$ (see more details in \S  3). Here  $t_{\ast}$ is related to the diffusion time scale in the disk $t_0$. 
Fore example, $t_{\ast}=4t_0/\pi^2$  for the linear dependence of viscosity $\hat\nu$ vs radius, 
$\psi=2$  [see an exponent of the  first term in series (\ref{y_signal0})].  

In this case we can calculate exactly the integral
\begin{equation}
\int_0^{\infty}Y^2(t)dt=\frac{1}{2t_{\ast}}
\label{Y2_expA}
\end{equation}
and then from Parseval 's theorem of Fourier transform we obtain that
\begin{equation}
\int_0^{\infty} ||F_{Y}(\omega)||^2d\omega=\int_0^{\infty}Y^2(t)dt=\frac{1}{2t_{\ast}}.
\label{FY_expA}
\end{equation}
In the general case of the spatial  distribution 
this can be estimated as follows
\begin{equation}
\int_0^{\infty}Y^2(t)dt\geq  [\int_0^{\infty}Y(t)dt]^2/\int_0^{{\hat D}t_{\ast}}dt \geq
\frac{1}{{\hat D}t_{\ast}}=\frac{1}{{D}t_{0}}.
\label{Y2_generalA}
\end{equation}
using H\"{o}lder's inequality. Here ${\hat D}\gax 2$ (compare with Eq. \ref{Y2_expA}). 
To derive inequality (\ref{Y2_generalA})  we also use the asymptotic behavior of $Y(t)$, namely 
$Y(t)\propto\exp(-t/t_{\ast}) $
 for $t\gg t_{\ast}$ (see details in \S 3).
Because 
$t_{\ast}\approx t_0/2.5$ (see Eq. \ref{y_signal0}) we can replace a product ${\hat D}t_{\ast}$ in 
Eq. (\ref{Y2_generalA}) by $Dt_0$ where $D\gax1$.
Then we find that
\begin{equation}
\int_0^{\infty} ||F_{Y}(\omega)||^2d\omega=\int_0^{\infty}Y^2(t)dt\geq\frac{1}{{D}t_{0}}.
\label{FY_generalA}
\end{equation} 
Finally Eqs (\ref{pwsp_varphi_res}), (\ref{FY_generalA}) gives  us the integrated total power of the resulting signal
\begin{equation}
P_x=\int_0^{\infty} ||F_{x}(\omega)||^2d\omega\sim\frac{1}{D{\cal Q}} \frac{P_{dr}}{\omega_{dr}t_{0}}
\label{power_LA}
\end{equation}
where ${\cal Q}=\omega_{dr}/\Delta\omega\gax 1$ is a quality of factor, $\Delta\omega$ is a HMFW of
$||F_{x}(\omega)||^2$.
To obtain equation (\ref{power_LA}) we also use  the mean value theorem.
\begin{equation}
\int_0^{\infty}||F_{dr}(\omega)||^2 ||F_{Y}(\omega)||^2d\omega=||F_{dr}(\omega_{\ast}||^2
\int_0^{\infty} ||F_{Y}(\omega)||^2d\omega
\label{FX_mean_prA}
\end{equation}
where 
\begin{equation}
||F_{dr}(\omega_{\ast}||^2\sim P_{dr}/({\cal Q}\omega_{dr}).
\label{FX_meanA}
\end{equation}
   
\section{Power spectrum. General treatment}

The Fourier transform of $Y(t)$ is calculated as follows (see e.g. Eq. \ref{ftg}, and Eqs. 
\ref{y_signal0}, \ref{Y_signal_general}):
\begin{equation}
F_{Y}(\omega)= \sum_{k=1}^{\infty}\frac{A_{k}}{i\omega + \lambda_{k}^{-2}}.
\label{ftgthick}
\end{equation}

Then the power spectrum is 
\begin{equation}
||F_{Y}(\omega)||^2=F_{Y}(\omega)\overline{F_{Y}(\omega)}=\sum_{k=1}^{\infty}\frac{A_{k}}{i\omega + \lambda_{k}^{-2}}\sum_{m=1}^{\infty}\frac{A_{m}}{-i\omega + \lambda_{k}^{-2}}.
\label{ftgpowerthick}
\end{equation}
Because for any complex values $a$ and $b$, 
\begin{equation}
a{\bar a} +b\bar b=|a|^2+|b|^2\geq (a{\bar b}+{\bar a}b) 
\label{(ab)}
\end{equation}
[where in our case $a=A_{k}/(i\omega + \lambda_k^{-2})$ and $b=A_{m}/(i\omega + \lambda_k^{-2})$] 
we can evaluate $||F_{Y}(\omega)||^2$ using Eq. (\ref{ftgpowerthick}) as follows
\begin{equation}
||F_{Y}(\omega)||^2\leq 2\sum_{k=1}^{\infty}\frac{A_{k}^2}{\omega^2 + \lambda_{k}^{-4}}.
\label{sc_appr_power}
\end{equation}

\section{Fourier transform of the FRED type fluctuation  $Y(t)$}

In section \S 3.3 we introduce the FRED type of the signal  $Y(t)$ (see Eq. \ref{FRED_gt}). 
The Fourier transforms of $Y(t)$ in this case is expressed through
the modified Bessel functions see PBM81, 2.3.16):
\begin{equation}
F_Y(\omega)\propto(C_0I_0+C_1I_1),
\label{gt_ftransform_bes}
\end{equation}
where
\begin{equation}
I_j=\left(\frac{\pi}{2}\right)^{1/2}\frac{2^{1-\alpha_j}t_0^{\alpha_j}\exp[-i(\alpha_j+1/2)\varphi/2]}{\rho^{\alpha_j/2+1/4}}\exp[-\rho^{1/2}
\cos(\varphi/2)-i\rho^{1/2}\sin(\varphi/2)]~~~{\rm for}~~j=0,~1
\label{I_j_integral}
\end{equation} 
where
$\alpha_0=1$ and $\alpha_1=1/2-\gamma$,
\begin{equation}
\rho=(z_1^4+ \omega^2 t_0^2)^{1/2},
\label{modul_rho_app}
\end{equation}
\begin{equation}
\varphi=\arcsin(\omega t_0/\rho).
\end{equation}
To present integrals $I_j$ by formula (\ref{I_j_integral}) we approximate the modified Bessel functions [e.g. Abramovitz \& Stegun (1970)]
\[K_{\alpha_j}(w)\approx \left(\frac{\pi}{2}\right)^{1/2}\exp(-w)/w^{1/2},\]
where 
\[w=\rho^{1/2}[\cos(\varphi/2)+i\sin(\varphi/2)],\]
because $|w|$ in  $K_{\alpha_j}(w)$ is always greater than 1. In fact, W01 showed that $z_1>1$ but $w\gax z_1^2$.
\section{The Eddington luminosity limit in the disk and mass outflow in the wind} 
The critical $\dot M_d^{crit}$ to launch the wind at a given disk radius $R$ is determined by an equality of the radiation pressure and gravitational force in a given annulus between $R$ and $R+dR$, namely 
\begin{equation}
\frac{\sigma(R)Q_d^{crit}(R)}{c}=m_p\left(\frac{H}{R}\right)\frac{GM}{R^3}.
\label{mdot_crit1}
\end{equation}
Here 
\begin{equation}
Q_d^{crit}(R)=\frac{3}{8\pi}\frac{GM\dot M_d^{crit}}{ R^3}[1-(R_{\ast}/R)^{1/2}]
\label{mdot_crit2}
\end{equation} 
is the disk  luminosity per cm$^2$ in the annulus between $R$ and $R+dR$ (see SS73), 
$R_{\ast}$ is the central object radius, that is 3 Schwarzchild radii for BH and the NS radius for NS;
$\sigma(R)$ is an effective plasma cross section which equals to  the electron Thomson cross-section $\sigma_{\rm T}$ if the disk plasma is fully ionized, otherwise $\sigma(R)>\sigma_{\rm T}$ (see e.g. Proga 2005); H is a geometrical half-thickness of the disk; M is a mass of the central object, BH mass or NS mass for BH and NS cases respectively.

Thus we obtain using Eqs (\ref{mdot_crit1}-\ref{mdot_crit2}) that 
\begin{equation}
\dot M_d^{crit}(R)=\frac{8\pi}{3}R \left(\frac{H}{R}\right)\frac{m_pc}{\sigma(R)}\frac{1}{[1-(R_{\ast}/R)^{1/2}]}.
\label{mdot_crit3}
\end{equation}
If $H/R$ is  constant through the disk (see SS73) then $\dot M_d^{crit}\propto R/\sigma(R)$.
In this case the critical  mass accretion rate $\dot M_{out,d}^{crit}$  at the outer disk radius $R_{out}$  is much higher  than that $\dot M_{in,d}^{crit}$ at the innermost  disk radius $R_{in}$, namely   
\begin{equation}
\frac{\dot M_{out,d}^{crit}}{M_{in,d}^{crit}}=\frac{R_{out}}{R_{in}}\frac{\sigma_{\rm T}}{\sigma(R_{out})}\frac{[1-(R_{\ast}/R_{in})^{1/2}]}{[1-(R_{\ast}/R_{out})^{1/2}]}.
\label{ratio}
\end{equation}
The disk works as a filter that does not allow to supply $\dot M$ higher than
$M_d^{crit}(R_{in})$ to the innermost part of the disk. Much larger fraction of the total mass inflow  is converted to the mass outflow when the mass supply to the disk from the companion is very high. This  likely  occurs in Cyg X-1, Cyg X-2 and Sco X-1 cases. 

The next question which should be addressed is what is the total luminosity of the disk $L_d^{crit}$ if at any annulus the disk emits at the Eddington limit (see Eq. \ref{mdot_crit3})? In order to answer to this question we should integrate  $Q_d^{crit}$ over the radius, namely 
\begin{equation}
L_{d,tot}^{crit}=\int_{R_{in}}^{R_{out}}Q_d^{crit}(R)4\pi RdR= 4\pi GM\left(\frac{H}{R}\right) \frac{m_pc}{\sigma_{\rm T}}\int_{R_{in}}^{R_{out}}\frac{\sigma_{\rm T}}{\sigma(R)}\frac{dR}{R}.
\label{lum_crit_tot_1}
\end{equation}
Because
\begin{equation}
\int_{R_{in}}^{R_{out}}\frac{\sigma_{\rm T}}{\sigma(R)}\frac{dR}{R}\approx 
\frac{\sigma_{\rm T}}{\sigma(R_{out})}\ln(R_{out}/R_{in})
\label{integral}
\end{equation}
and by definition
\begin{equation}
L_{\rm Edd}=\frac{4\pi GM m_p c}{\sigma_{\rm T}}
\label{lum_edd}
\end{equation}
we have that
\begin{equation}
L_{d,tot}^{crit}=\left[\left(\frac{H}{R}\right)\frac{\sigma_{\rm T}}{\sigma(R_{out})}\ln(R_{out}/R_{in})\right]L_{\rm Edd}.
\label{lum_crit_tot_2}
\end{equation}
On the other hand using Eqs. (\ref{mdot_crit2}, \ref{mdot_crit3})  we obtain that the disk luminosity $L_{d,in}^{crit}$ with the constant $\dot M_{in,d}^{crit}$  is
\begin{equation}
L_{d,in}^{crit}=\int_{R_{in}}^{R_{out}}Q_d^{crit}(R_{in})4\pi RdR\sim\left(\frac{H}{R}\right)L_{\rm Edd}.
\label{lum_crit_in}
\end{equation}
Thus we expect the saturation of  the total disk luminosity   with the mass supply to the disk because 
\begin{equation}
\frac{L_{d,tot}^{crit}}{L_{d,in}^{crit}}\sim 2\times \left[\frac{\ln(R_{out}/R_{in})}{10}\right]
\left[\frac{\sigma_{\rm T}/\sigma(R_{out})}{0.2}\right].
\label{lum_ratio}
\end{equation}
Here we use representative  values of the ratios 
$R_{out}/R_{in}\sim 10^4$ (see e.g. SS73) and $\sigma_{\rm T}/\sigma(R_{out})\sim0.2$ (Proga 2005).

Now we can calculate the total mass outflow rate in the sources in which mass accretion supply from companion is very high, i.e about  $\dot M_d^{crit}(R_{out})$.  Because 
$\dot M_d^{crit}(R)<$ $\dot M_d^{crit}(R+dR)$
(see Eq. \ref{mdot_crit3})  the surplus  ( overflow) $\Delta \dot M_d^{crit}(R)$   in the annulus 
$(R, ~R+dR)$ emerges  as outflow there, i.e. the mass outflow rate is 
$\Delta \dot M_{outflow}^{crit}(R)=\Delta \dot M_d^{crit}(R)$.   Consequently, the total mass outflow rate from the disk is 
\begin{equation}
\dot M_{outflow}^{crit}=\int_{R_{in}}^{R_{out}}\Delta \dot M_d^{crit}(R)dR\sim \dot M_d^{crit}(R_{out})
\label{outflow}
\end{equation}
which is much higher than the critical mass accretion rate at inner disk radius
\begin{equation}
\frac{\dot M_{outflow}^{crit}}{\dot M_d^{crit}(R_{in})}\sim \frac{R_{out}}{R_{in}}\gg 1.
\label{outflow_ratio}
\end{equation}
The Thomson optical depth of outflow $\tau_{W}$ with  a constant velocity $v$  can be calculated  using the continuity equation
\begin{equation}
 M_{outflow}^{crit}=4\pi R^2n\sigma_{\rm T}m_pv
\label{cont_eq}
\end{equation}
 and 
Eq. (\ref{outflow}), namely
\begin{equation}
\tau_{W}=\int_{R_{out}}^{\infty}n\sigma_T dR=\frac{\dot M_{outflow}^{crit}\sigma_{\rm T}}{4\pi m_p v}\int_{R_{out}}^{\infty}
\frac{dR}{R^{2}}.
\label{tau outflow1}
\end{equation}
Finally we obtain using Eqs. (\ref{mdot_crit3}), (\ref{outflow}) and (\ref{tau outflow1}) that
\begin{equation}
\tau_{W}=\frac{\dot M_{outflow}^{crit}\sigma_{\rm T}}{4\pi m_p v R_{out}}=\frac{2}{3}\frac{H}{R}
\frac{\sigma_{\rm T}}{\sigma(R)}\frac{c}{v}
\label{tau outflow2}
\end{equation}
and thus $\tau_W\gax 1$ for typical values of $H/R\sim 0.1$, $\sigma_{\rm T}/\sigma(R)\sim 0.2$ and 
$v/c\sim (R_{in}/R_{out})^{1/2}\sim 0.01$. To evaluate the wind velocity $v$ we use a formula for the wind terminal velocity derived  in Rybicki \& Lightman (1979) (see Chapter 1 there) and the value of
$R_{out}/R_{in}\sim 10^{4}$. 

\newpage

\newpage

\newpage

\newpage

\newpage

\newpage

\newpage

\newpage

\newpage

\newpage

\newpage

\newpage
\newpage

\begin{landscape}
\begin{deluxetable}{lllllllll}
\tabletypesize{\footnotesize}
\tablewidth{0pt}
\tablecaption{Summary of RXTE data}
\tablehead{\colhead{Observation ID(\tablenotemark{a})} & \colhead{Date, MJD} & 
\colhead{$\psi_{CC}$} & \colhead{$t_{C,0}$, s}& \colhead{$\psi_{D}$}& 
\colhead{$t_{*}$, ms}&\colhead{$\Gamma$}&\colhead{$\nu_{QPO}$, Hz}&\colhead{$\chi^2/N_{dof}$}}
\startdata
10257-01-01-00 & 50242.1 & $ ... $ & $ ... $ & $ 3.031\pm 0.016 $ & $  41(3) $ & $1.017_{-0.018}^{+0.030}$ & - & 162.0/220 \\
30158-01-03-00 & 50796.4 & $ 2.531\pm 0.070 $ & $ 6.284\pm 1.375 $ & $ ... $ & $  13(2) $ & $0.513_{-0.002}^{+0.005}$ & $0.355_{-0.095}^{+0.166}$ & 96.3/185 \\
40099-01-03-01 & 51220.3 & $ 2.324\pm 0.119 $ & $ 1.793\pm 0.321 $ & $ ... $ & $   9(3) $ & $0.550_{-0.003}^{+0.006}$ & $1.240_{-0.084}^{+0.144}$ & 114.7/213 \\
40099-01-05-00 & 51251.2 & $ 2.067\pm 0.073 $ & $ 1.236\pm 0.116 $ & $ ... $ & $   8(2) $ & $0.648_{-0.002}^{+0.005}$ & $1.579_{-0.128}^{+0.246}$ & 178.7/216 \\
40099-01-19-00 & 51446.7 & $ 1.696\pm 0.093 $ & $ 0.352\pm 0.026 $ & $ ... $ & $   5(2) $ & $0.778_{-0.010}^{+0.012}$ & $4.205_{-0.109}^{+0.213}$ & 147.1/216 \\
40099-01-21-00 & 51475.7 & $ 2.170\pm 0.058 $ & $ 0.644\pm 0.055 $ & $ ... $ & $   6(2) $ & $0.703_{-0.004}^{+0.007}$ & $3.318_{-0.105}^{+0.208}$ & 157.6/216 \\
50109-01-01-00 & 51900.4 & $ -0.753\pm 0.002 $ & $ 0.034\pm 0.001 $ & $ 2.736\pm 0.076 $ & $  22(2) $ & $1.035_{-0.003}^{+0.005}$ & - & 219.6/217 \\
50109-01-02-03 & 51904.2 & $ 0.105\pm 0.562 $ & $ 0.060\pm 0.013 $ & $ ... $ & $ ... $ & $0.883_{-0.006}^{+0.011}$ & $7.083_{-0.234}^{+0.446}$ & 153.8/216 \\
50109-01-05-01 & 51934.0 & $ 0.438\pm 0.246 $ & $ 0.061\pm 0.006 $ & $ 1.438\pm 1.062 $ & $   2(2) $ & $0.927_{-0.003}^{+0.006}$ & $8.703_{-0.106}^{+0.207}$ & 383.5/214 \\
50110-01-52-00 & 52385.3 & $ ... $ & $ ... $ & $ 2.937\pm 0.015 $ & $ 106(5) $ & $1.558_{-0.011}^{+0.023}$ & - & 252.1/220 \\
50119-01-01-00 & 51845.6 & $ 0.853\pm 0.003 $ & $ 0.045\pm 0.001 $ & $ 2.654\pm 0.106 $ & $  22(1) $ & $0.979_{-0.006}^{+0.012}$ & - & 307.9/189\\
50119-01-03-02 & 51851.7 & $ -0.835\pm 0.204 $ & $ 0.038\pm 0.002 $ & $ 1.928\pm 0.805 $ & $ ... $ & $0.950_{-0.004}^{+0.007}$ & $8.363_{-0.173}^{+0.339}$ & 207.0/186\\
50119-01-04-01 & 51897.1 & $ 1.632\pm 0.105 $ & $ 0.255\pm 0.023 $ & $ ... $ & $   3(2) $ & $0.810_{-0.002}^{+0.007}$ & $4.122_{-0.152}^{+0.294}$ & 188.1/213\\
60089-02-01-00 & 52175.7 & $ ... $ & $ ... $ & $ 3.133\pm 0.011 $ & $  52(2) $ & $1.169_{-0.004}^{+0.010}$ & - & 156.4/220\\
60089-02-01-01 & 52175.8 & $ 1.696\pm 0.735 $ & $ 0.026\pm 0.009 $ & $ 3.078\pm 0.021 $ & $  42(4) $ & $1.138_{-0.005}^{+0.008}$ & - & 199.6/217\\
60089-02-02-00 & 52210.7 & $ 1.537\pm 0.107 $ & $ 0.214\pm 0.019 $ & $ ... $ & $   2(1) $ & $0.807_{-0.005}^{+0.008}$ & $5.132_{-0.149}^{+0.278}$ & 186.3/216\\
60090-01-01-00 & 52341.2 & $ ... $ & $ 0.052\pm 0.001 $ & $ 2.785\pm 0.051 $ & $  30(1) $ & $1.043_{-0.003}^{+0.008}$ & - & 256.5/217\\
60090-01-02-00 & 52358.2 & $ ... $ & $ ... $ & $ 2.801\pm 0.016 $ & $  52(2) $ & $1.286_{-0.003}^{+0.007}$ & - & 213.3/220\\
60090-01-06-00 & 52415.1 & $ ... $ & $ ... $ & $ 2.960\pm 0.013 $ & $  48(2) $ & $1.248_{-0.005}^{+0.013}$ & - & 154.1/220\\
60090-01-07-03 & 52426.3 & $ ... $ & $ ... $ & $ 3.260\pm 0.014 $ & $  64(4) $ & $1.112_{-0.004}^{+0.008}$ & - & 138.3/220\\
60090-01-11-01 & 52481.9 & $ ... $ & $ ... $ & $ 2.845\pm 0.021 $ & $  90(6) $ & $1.575_{-0.009}^{+0.025}$ & - & 172.3/220\\
60090-01-12-01 & 52497.9 & $ ... $ & $ ... $ & $ 2.894\pm 0.015 $ & $  72(4) $ & $1.493_{-0.009}^{+0.022}$ & - & 235.4/220\\
60090-01-12-02 & 52498.0 & $ ... $ & $ ... $ & $ 3.001\pm 0.021 $ & $ 112(9) $ & $1.415_{-0.012}^{+0.031}$ & - & 173.1/220\\
60090-01-14-01 & 52524.8 & $ ... $ & $ ... $ & $ 2.967\pm 0.014 $ & $  57(3) $ & $1.339_{-0.011}^{+0.017}$ & - & 203.0/220\\
60090-01-14-02 & 52524.9 & $ ... $ & $ ... $ & $ 2.931\pm 0.016 $ & $  52(3) $ & $1.298_{-0.006}^{+0.017}$ & - & 152.7/220\\
60090-01-14-04 & 52525.0 & $ ... $ & $ ... $ & $ 3.036\pm 0.014 $ & $  61(3) $ & $1.394_{-0.008}^{+0.016}$ & - & 153.8/220\\
60090-01-15-01 & 52537.7 & $ ... $ & $ ... $ & $ 3.073\pm 0.010 $ & $  41(2) $ & $1.204_{-0.006}^{+0.013}$ & - & 149.6/220\\
60090-01-16-00 & 52553.7 & $ -0.238\pm 0.228 $ & $ 0.050\pm 0.004 $ & $ ... $ & $   6(2) $ & $0.909_{-0.003}^{+0.006}$ & $6.308_{-0.293}^{+0.589}$ & 241.0/215\\
60136-03-01-00 & 52088.2 & $ 2.073\pm 0.080 $ & $ 1.098\pm 0.130 $ & $ ... $ & $ ... $ & $0.583_{-0.004}^{+0.007}$ & $2.096_{-0.216}^{+0.407}$ & 110.4/217\\
80110-01-12-00 & 53241.8 & $ 2.246\pm 0.063 $ & $ 0.914\pm 0.087 $ & $ ... $ & $   7(2) $ & $0.645_{-0.004}^{+0.007}$ & $2.434_{-0.178}^{+0.344}$ & 133.4/216\\
80110-01-13-00 & 53251.7 & $ 2.026\pm 0.041 $ & $ 0.547\pm 0.008 $ & $ ... $ & $   4(2) $ & $0.729_{-0.003}^{+0.005}$ & $3.955_{-0.112}^{+0.220}$ & 150.8/216 \\
80110-01-14-00 & 53267.7 & $ ... $ & $ ... $ & $ 3.073\pm 0.009 $ & $  49(2) $ & $1.081_{-0.004}^{+0.014}$ & - & 203.1/220 \\
80110-01-14-03 & 53267.9 & $ ... $ & $ ... $ & $ 3.151\pm 0.014 $ & $  49(3) $ & $1.107_{-0.004}^{+0.007}$ & - & 136.6/220 \\
80110-01-15-00 & 53279.7 & $ ... $ & $ 0.027\pm 0.001 $ & $ 2.961\pm 0.039 $ & $  32(3) $ & $1.071_{-0.007}^{+0.008}$ & - & 238.6/218 \\
80110-01-16-00 & 53293.7 & $ -0.351\pm 0.565 $ & $ 0.062\pm 0.012 $ & $ ... $ & $ ... $ & $0.904_{-0.004}^{+0.006}$ & $6.687_{-0.230}^{+0.444}$ & 167.6/215 \\
80110-01-18-00 & 53321.6 & $ -1.171\pm 0.175 $ & $ 0.128\pm 0.002 $ & $ 3.414\pm 0.322 $ & $  14(20) $ & $0.729_{-0.004}^{+0.008}$ & $4.287_{-0.134}^{+0.263}$ & 142.3/214 \\
80110-01-20-04 & 53352.7 & $ 0.825\pm 0.338 $ & $ 0.143\pm 0.022 $ & $ ... $ & $ ... $ & $0.862_{-0.004}^{+0.007}$ & $5.373_{-0.171}^{+0.334}$ & 166.7/215 \\
80110-01-22-00 & 53378.5 & $ ... $ & $ ... $ & $ 3.180\pm 0.014 $ & $  61(4) $ & $1.249_{-0.016}^{+0.020}$ & - & 131.1/220 \\
80110-01-23-00 & 53391.4 & $ 1.414\pm 0.126 $ & $ 0.201\pm 0.018 $ & $ ... $ & $ ... $ & $0.862_{-0.005}^{+0.007}$ & $5.568_{-0.116}^{+0.228}$ & 188.2/216 \\
80110-01-23-04 & 53391.7 & $ -0.295\pm 0.499 $ & $ 0.067\pm 0.011 $ & $ ... $ & $ ... $ & $0.901_{-0.005}^{+0.008}$ & $6.993_{-0.209}^{+0.404}$ & 181.3/216 \\

\enddata
\end{deluxetable}
\end{landscape}

\end{document}